\documentclass[twocolumn,aps,pr,superscriptaddress,preprintnumbers,nofootinbib,10pt]{revtex4-2} 

\usepackage{amsmath,amssymb}
\usepackage[dvipdf,dvips]{graphicx}
\usepackage{color}
\usepackage{hyperref}
\usepackage{url}
\usepackage{slashed}
\usepackage{subfigure}
\usepackage[usenames,dvipsnames]{xcolor}
\usepackage{amsmath}
\usepackage{amsfonts}
\usepackage{float} 
\usepackage{amssymb}
\usepackage{epsfig}
\usepackage{graphics}
\usepackage{euscript}
\usepackage{slashed}
\usepackage{epstopdf}
\usepackage[utf8]{inputenc}
\allowdisplaybreaks
\usepackage[normalem]{ulem}
\usepackage{pifont}
\usepackage{dsfont}
\usepackage{MnSymbol}
\usepackage{verbatim}
\usepackage{graphicx}
\usepackage{latexsym}

\usepackage{yfonts}
\usepackage{MnSymbol}

 \usepackage{feynmp-auto}
 \usepackage[compat=1.0.0]{tikz-feynman}

\hypersetup{
colorlinks=true,
citecolor=blue,
citebordercolor=red,
linktoc=all,
linkcolor=blue,
urlcolor=blue
}

\graphicspath{{./figs/}}
\newbox{\ORCIDicon}
\sbox{\ORCIDicon}{\large
                  \includegraphics[width=0.8em]{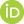}}

\begin{document}

\title{Yang-Mills theories at finite temperature quantized in linear covariant gauges: gauge copies and semi-non-perturbative effects}

\author{Luigi C. Ferreira\,\href{https://orcid.org/0000-0003-1467-4260}{\usebox{\ORCIDicon}}}
\email{luigi\_carvalho@id.uff.br}
\affiliation{Universidade do Estado do Rio de Janeiro, R. S\~ao Francisco Xavier 524, 20550-013, Maracan\~a, Rio de Janeiro, RJ, Brazil}



\begin{abstract}
We consider four-dimensional Euclidean Yang-Mills theories quantized in the maximal Abelian and linear covariant gauges at finite temperature. Non-perturbatively, the Faddeev-Popov procedure must be improved to take into account the existence of the so-called Gribov copies. Tapping on previous results about the elimination of infinitesimal Gribov copies in maximal Abelian and linear covariant gauges at zero temperature, we explore the interplay between finite temperature effects and the removal of gauge copies. We focus in a hybrid approach where the thermal masses are derived through perturbative propagators as a stepping stone for a self-consistent treatment. The resulting action collects the effects of the elimination of infinitesimal Gribov copies as well as the thermal masses. We verify the existence of three different phases for the gluonic degrees of freedom; one of complete confinement at low temperatures, an intermediate one of  partial confinement, and one of complete deconfinement at high temperatures. 
\end{abstract}


\maketitle

\section{Introduction}
\label{intro}
The treatment of Yang-Mills theories by means of continuum quantum-field theoretic tools requires the introduction of a gauge-fixing term which engenders a well-defined gauge field propagator. This is at the core of the evaluation of $n$-point correlation functions, either by a perturbative scheme or non-perturbative methods, such as functional equations \cite{vasiliev2019functional,aguilar2023schwinger, maris2003dyson,fu2022qcd,alkofer2019bound}. In a functional integral language, the Faddeev-Popov method is the standard textbook prescription to fix a gauge \cite{faddeev1967feynman,zwanziger2004nonperturbative,ryder1996quantum}. In short, one introduces an identity in the functional measure which is composed of a gauge-fixing condition and a Jacobian which stands for the so-called Faddeev-Popov determinant. In perturbation theory, i.e., when the product of the gauge coupling and the gauge field is sufficiently small, the hypothesis of the Faddeev-Popov procedure are well justified and, indeed, the method is suitable to select one representative of the gauge field per gauge orbit. However, as pointed out in \cite{gribov1978quantization} by Gribov and extended in \cite{zwanziger1981covariant,zwanziger1989local}, this is not true beyond perturbation theory. In fact, towards strongly-coupled regimed, the assumptions behind the Faddeev-Popov procedure are not satisfied and this can be explicitly checked in different gauges, see, e.g., \cite{zwanziger1989local,sobreiro2005study,capri2009study,capri2016more,capri2015gribov}. There are normalizable configurations which satisfy the gauge condition and belong to the same gauge orbit. Hence, the gauge-fixing condition does not have a unique solution per gauge orbit, spoiling one of the assumptions of the Faddeev-Popov procedure. Those spurious configurations are known as Gribov (or simply gauge) copies. 

One possible strategy to deal with the existence of Gribov copies was initiated by Gribov \cite{gribov1978quantization} and pushed forward by Zwanziger \cite{zwanziger1981covariant,zwanziger1989local}. The main idea consists in fixing the gauge (in their case, the Landau gauge) using the standard Faddeev-Popov procedure and on top of that, the imposition of a restriction of the functional integral to a region free of Gribov copies, the so-called Gribov region. Such a region has very important geometrical properties as proved in \cite{dell1991every,vandersickel2012gribov,cucchieri2013crossing,maas2016more,vandersickel2011study} and, most importantly, all gauge orbits cross it at least once. This means that such a restriction does not leave out any ``physical configuration". Nonetheless, the Gribov region is just free of the so-called infinitesimal Gribov copies, i.e., those generated by infinitesimal gauge transformations. Finite copies are still present in the Gribov region \cite{vandersickel2012gribov,dell1991every,van1992more} and a truly free-of-copies domain is the so-called Fundamental Modular Region (FMR). Unfortunately, there is no practical method to date that implements the restriction of the functional integral to the FMR. Much of the progress just reported was due to particular properties of the Landau gauge and it is not true that those properties are valid in different gauges. Consequently, dealing with the Gribov problem in different gauges is a challenging problem. Yet, the choice of gauge should not impact the correlation functions of observables and, thus, any proposed solution in a given gauge should not ignore such a feature. In the Landau gauge, the restriction of the path integral to the Gribov region led to the so-called Gribov-Zwanziger (GZ) action which was further refined by the introduction of dimension-two condensates leading to the so-called Refined Gribov-Zwanziger (RGZ) \cite{dudal2008refinement,dudal2008new}. For reviews on the topic, we refer to, e.g., \cite{vandersickel2011study}.

In this context, our work delves into the RGZ model at finite temperature. The primary focus is to investigate the impact of the Gribov parameter and the masses of the condensates on the pole structure of the propagators of the theory and their behavior, particularly in relation to phase transitions driven by variations in temperature, denoted as $T$. This exploration is carried out using a semi-classical technique previously followed by Canfora in \cite{canfora2014gribov}, which involves introducing an effective massive term to the Yang-Mills theory to account for thermal effects in the gluon propagator. Specifically, we incorporate a term $\frac{1}{2}M^{2}A_{\mu}A_{\mu}$ within the theoretical framework's action, with $M^{2}$ representing $1$-loop finite-temperature corrections and thermal information.

The presence of this thermal mass plays a pivotal role in the emergence of a deconfined regime. Through our analysis, we identify two critical temperatures associated with phase transitions. The first transition involves shifting from a fully confined phase to a partially confined one, characterized by the coexistence of physical and unphysical degrees of freedom. The second transition pertains to the shift from partial confinement to total deconfinement.

Our findings align with previous research in the Landau gauge \cite{canfora2014gribov}, and they extend the scope to encompass Linear covariant gauges and the impact of RGZ condensate masses. It is evident from our analysis that these condensate masses contribute to a reduction in phase transition temperatures within the theoretical framework.

The work is divided in the following sections: in section \ref{sec2} we implement the quantization of the Yang-Mills theory in the linear covariant gauges. In section \ref{sec 3} we obtain the gluon self-energy of the theory and we calculate the longitudinal and transverse thermal masses from it introducing also the semi-classical approach that we will use in order to reproduce these effects in the tree-level propagators of the theory. In section \ref{section4} we introduce the GZ panorama related to the elimination of the infinitesimal Gribov copies in the Landau gauge. In section \ref{extensionlc} we extend it to linear covariant gauges. In section \ref{section6} we introduce the refinement of the GZ approach, in which we show how to build an effective BRST-invariant action that capture the effects due to non-trivial condensates that appear due to instabilities of the GZ action in the infrared regime. In section \ref{section7} we derive our effective gluon propagators in the RGZ framework that take into account thermal effects. From \ref{section8} to \ref{section11} we make the analysis of the regimes of the theory and we calculate the critical temperatures of phase transition. In section \ref{section12} we extend the calculation of the thermal masses to the RGZ panorama at 1-loop and then we discuss the results and compare them with previous literature.

\section{Yang-Mills theory in the Linear covariant gauge}\label{sec2}
We start writing the path integral of the $SU(N)$ Yang-Mills theory, which is given by,
\begin{equation}
\EuScript{Z}=\int{[\EuScript{D}A]}\,{\rm e}^{-S_{\rm YM}}\,,
\label{z}
\end{equation}
where $S_{\rm YM}$ is the Yang-Mills action, given by
\begin{equation}
S_{\rm YM}=\frac{1}{4}\int{d^{4}x}F_{\mu\nu}^{a}F_{\mu\nu}^{a}.\label{s1}
\end{equation}
Using the Faddeev-Popov quantization procedure we find out that in the Landau gauge the path integral (\ref{z}) becomes,
\begin{eqnarray}
Z=\mathcal{N}\int{[DA][D\bar{c}][Dc][Db]}e^{-S_{YM}-S_{FP}^{L}}\label{zz},
\end{eqnarray}
where,
\begin{eqnarray}
S_{FP}^{L}=\int{d^{4}x}\Bigg(ib^{a}\partial_{\mu}A_{\mu}^{a}+\bar{c}^{a}\partial_{\mu}D_{\mu}^{ab}c^{b}\Bigg)\label{fplandau}.
\end{eqnarray}
Doing so, means that we are implementing the following gauge fixing condition,
\begin{eqnarray}
\partial_{\mu}A_{\mu}^{a}=0.
\end{eqnarray}
However, the Landau gauge is just a special case of a more general class of gauge fixing conditions called 'Linear covariant gauges'. This gauges are given by,
\begin{eqnarray}
\partial_{\mu}A_{\mu}^{a}=i\alpha b^{a},\label{gauge LC}
\end{eqnarray}
from where it can easily be seen that the Landau gauge is recovered in the limit where $\alpha\rightarrow0$.

In order to properly implement the condition (\ref{gauge LC}) into the path integral (\ref{zz}), we need to add a term $\frac{\alpha}{2}b^{a}b^{a}$ into the Faddeev-Popov action (\ref{fplandau}), in such a way that,
\begin{eqnarray}
S_{FP}^{LC}&=&\int{d^{4}x}\Bigg(ib^{a}\partial_{\mu}A_{\mu}^{a}+\frac{\alpha}{2}b^{a}b^{a}-\bar{c}^{a}M^{ab}c^{b}\Bigg),\label{fplc1}
\end{eqnarray}
with,
\begin{eqnarray}
M^{ab}=-\partial_{\mu}D_{\mu}^{ab},\label{FP operator}
\end{eqnarray}
being the Faddeev-Popov operator, and,
\begin{eqnarray}
D_{\mu}^{ab}=\delta^{ab}\partial_{\mu}-gf^{abc}A_{\mu}^{c},
\end{eqnarray}
being the covariant derivative in the adjoint representation.
So the total action becomes,
\begin{eqnarray}
S=S_{YM}+S_{FP}^{LC},\label{actionLC}
\end{eqnarray}
and the path integral for the Yang-Mills theory in the linear covariant gauges turns out to be,
\begin{eqnarray}
Z=\mathcal{N}\int{[DA][D\bar{c}][Dc][Db]}e^{-S}.\label{z1}
\end{eqnarray}

\section{The gluon self-energy and the thermal mass}\label{sec 3}

In this section we will calculate the thermal contribution to the poles of the longitudinal and transverse sectors of the gluon propagator. This will be achieved by means of the components $\Pi_{44}^{(2)}$ and $\Pi_{\mu\mu}^{(2)}$ of the gluon self-energy $\Pi_{\mu\nu}^{(2)}$.

In order to proceed, it is necessary to initially write down the propagators and vertices of the theory, so enabling the construction of diagrams that will contribute to the gluon self-energy. 


From eq.(\ref{actionLC}) we obtain the gluon and ghost propagators in the Euclidean space, given respectively by,
\begin{eqnarray}
\langle A_{\mu}^{a}(k)A_{\nu}^{b}(-k)\rangle&=&\Bigg[\frac{1}{k^{2}}\Bigg(g_{\mu\nu}-\frac{k_{\mu}k_{\nu}}{k^{2}}\Bigg)+\alpha\frac{k_{\mu}k_{\nu}}{k^{4}}\Bigg]\delta^{ab},\nonumber\\
\langle \bar{c}^{a}(k)c^{b}(-k)\rangle&=&-\frac{\delta^{ab}}{k^{2}}.
\end{eqnarray}
Concerning the vertices, we can also easily see that they are given by,
\begin{eqnarray}
\left[V_{A^{3}}(q_{1},q_{2},q_{3})\right]_{\alpha\beta\lambda}^{gfh}&=&(2\pi)^{4}igf^{gfh}\Bigg((q_{1}-q_{3})_{\beta}\delta_{\alpha\lambda}+\nonumber\\
&+&(q_{3}-q_{2})_{\alpha}\delta_{\beta\lambda}+(q_{2}-q_{1})_{\lambda}\delta_{\alpha\beta}\Bigg),\nonumber\\
\left[V_{A^{4}}(q_{1},q_{2},q_{3},q_{4})\right]_{\alpha\beta\gamma\sigma}^{gfhn}&=&
-(2\pi)^{4}g^{2}\times\nonumber\\
&\times&\Bigg(f^{abe}f^{cde}(\delta_{\alpha\gamma}\delta_{\sigma\beta}-\delta_{\sigma\alpha}\delta_{\beta\gamma})+\nonumber\\
&+&f^{ade}f^{bce}(\delta_{\alpha\beta}\delta_{\sigma\gamma}-\delta_{\alpha\gamma}\delta_{\sigma\beta})+\nonumber\\
&+&f^{ace}f^{bde}(\delta_{\alpha\beta}\delta_{\sigma\gamma}-\delta_{\sigma\alpha}\delta_{\beta\gamma})\Bigg),\nonumber\\
\left[V_{A\bar{c}c}(k,p,r)\right]_{\mu}^{abc}&=&-(2\pi)^{4}igf^{abc}p_{\mu},\label{vert123}
\end{eqnarray}
where $p$ is the momentum of the $\bar{c}^{b}(p)$ field.

We can now build the diagrams that contribute to the gluon's self-energy. In order to accomplish this, we will contemplate a positive sign for all momenta entering the vertex and a negative sign for those leaving it when constructing the vertices. In addition, as we will be performing calculations in the Minkowski space-time, we will build the diagrams incorporating the Wick rotation $k\rightarrow ik$. With this in mind, we have that at $1$-loop they are given by,

\begin{tikzpicture}
		\begin{feynman}
			\vertex (a);
			\vertex [right=0.5cm of a] (b);
			\vertex [right=0.5cm of b] (c);
			\vertex [right=0.5cm of c] (d);
			\vertex [right=0.15cm of d] (e) {$=$};
			\vertex [right=0.25cm of e] (f);
			\vertex [right=0.5cm of f] (g);
			\vertex [right=0.5cm of g] (h);
			\vertex [right=0.5cm of h] (i);
			\vertex [right=0.15cm of i] (j) {+};
			\vertex [right=0.25cm of j] (k);
			\vertex [right=0.75cm of k] (l);
			\vertex [above=0.5cm of l] (m);
			\vertex [right=0.75cm of l] (n);
                \vertex [below=0.25cm of l] (l1) {$(b)$};
			\vertex [right=0.15cm of n] (o) {+};
			\vertex [right=0.25cm of o] (p);
			\vertex [right=0.5cm of p] (q);
			\vertex [right=0.5cm of q] (r);
			\vertex [right=0.5cm of r] (s);

			\diagram{
				(a) -- [boson] (b);
			    (b) --[ half left, fill=gray] (c);
				(c) --[ half left, fill=gray] (b);
				(c) -- [boson] (d);
			
				(f) -- [boson] (g);
				(g) --[ghost, half left] (h);
				(h) --[ghost, half left, edge label=$(a)$] (g);
				(h) -- [boson] (i);
			
				(k) -- [boson] (l);
				(l) --[boson, half left] (m);
				(m) --[boson, half left] (l);
				(l) -- [boson] (n);
			
				(p) -- [boson] (q);
				(q) --[boson, half left] (r);
				(r) --[boson, half left, edge label=$(c)$] (q);
				(r) -- [boson] (s);
				};
			
		\end{feynman}\label{aaaaa}

	\end{tikzpicture}

The diagram $(a)$, which is the ghost sunrise diagram, is given by,
\begin{eqnarray}
i\Pi_{\mu\nu}^{(2)ab}(k)&=&-g^{2}f^{acd}f^{bc'd'}\int{\frac{d^{d}p}{(2\pi)^{d}}}\Bigg(p_{\mu}\left(k+p\right)_{\nu}\times\nonumber\\
&\times&\left(-\frac{\delta^{cc'}}{p^{2}}\right)\left(-\frac{\delta^{dd'}}{(k+p)^{2}}\right)\Bigg)\nonumber\\
&=&-g^{2}f^{acd}f^{bcd}\int{\frac{d^{d}p}{(2\pi)^{d}}}\Bigg(\frac{p_{\mu}\left(k+p\right)_{\nu}}{p^{2}(k+p)^{2}}\Bigg)\nonumber\\
&=&-g^{2}N\delta^{ab}\int{\frac{d^{d}p}{(2\pi)^{d}}}\Bigg(\frac{p_{\mu}\left(k+p\right)_{\nu}}{p^{2}(k+p)^{2}}\Bigg),\label{Acc}
\end{eqnarray}
while the gluon tadpole of diagram $(b)$ is,
\begin{eqnarray}
i\Pi_{\mu\nu}^{(2)ab}(k)&=&\frac{1}{2}g^{2}\Bigg(f^{abe}f^{cqe}\left(\delta_{\mu\delta}\delta_{\eta\nu}-\delta_{\eta\mu}\delta_{\nu\delta}\right)+\nonumber\\
&+&f^{aqe}f^{bce}\left(\delta_{\mu\nu}\delta_{\eta\delta}-\delta_{\mu\delta}\delta_{\eta\nu}\right)+\nonumber\\
&+&f^{ace}f^{bqe}\left(\delta_{\mu\nu}\delta_{\eta\delta}-\delta_{\eta\mu}\delta_{\nu\delta}\right)\Bigg)\times\nonumber\\
&\times&\int{\frac{d^{d}p}{(2\pi)^{d}}}\Bigg[\frac{\delta^{cq}}{p^{2}}\left(\delta_{\delta\eta}-(1-\alpha)\frac{p_{\delta}p_{\eta}}{p^{2}}\right)\Bigg]\nonumber\\
&=&g^{2}N\delta^{ab}\int{\frac{d^{d}p}{(2\pi)^{d}}}\Bigg[\alpha\Bigg(\frac{-p_{\mu}p_{\nu}+(p\cdot p)\delta_{\mu\nu}}{p^{4}}\Bigg)+\nonumber\\
&+&\frac{p_{\mu}p_{\nu}+(d-2)(p\cdot p)\delta_{\mu\nu}}{p^{4}}\Bigg].\label{AAAA}
\end{eqnarray}
and the gluon sunrise diagram $(c)$ is,
\begin{eqnarray}
i\Pi_{\mu\nu}^{(2)ab}(k)&=&\frac{1}{2}g^{2}f^{acd}f^{c'bd'}\int{\frac{d^{d}p}{(2\pi)^{d}}}\Bigg[\Bigg((k-p)_{\tau}\delta_{\mu\lambda}+\nonumber\\
&+&(p-(-k-p))_{\mu}\delta_{\tau\lambda}+(-k-p-k)_{\lambda}\delta_{\mu\tau}\Bigg)\times\nonumber\\
&\times&\Bigg((k+p-(-p))_{\nu}\delta_{\rho\delta}+(-p-(-k))_{\rho}\delta_{\nu\delta}+\nonumber\\
&+&(-k-(k+p))_{\delta}\delta_{\rho\nu}\Bigg)\frac{\delta^{dd'}}{p^{2}}\left(\delta_{\lambda\delta}-(1-\alpha)\frac{p_{\lambda}p_{\delta}}{p^{2}}\right)\times\nonumber\\
&\times&\frac{\delta^{cc'}}{(k+p)^{2}}\left(\delta_{\tau\rho}-(1-\alpha)\frac{(k+p)_{\tau}(k+p)_{\rho}}{(k+p)^{2}}\right)\Bigg]\label{AAA}.\nonumber\\
\end{eqnarray}

By applying the transformations $t\rightarrow it$ and $p\rightarrow(p-k)$ and utilizing the approximation $p>>k$ in the numerator, the equation (\ref{Acc}) for the gluon-ghost diagram yields the following result,
%
\begin{eqnarray}
i\Pi_{\mu\nu}^{(2)ab}(k)=-g^{2}N\delta^{ab}\int{\frac{d^{d}p}{(2\pi)^{d}}}\Bigg(\frac{p_{\mu}p_{\nu}}{p^{2}(p-k)^{2}}\Bigg)\label{Acck0}
\end{eqnarray}
Repeating the same procedure on eq.(\ref{AAA}), we obtain that,
\begin{eqnarray}
i\Pi_{\mu\nu}^{(2)ab}(k)&=&g^{2}N\delta^{ab}\int{\frac{d^{d}p}{(2\pi)^{d}}}\Bigg[\frac{\alpha}{p^{4}}\left(p_{\mu}p_{\nu}-p^{2}\delta_{\mu\nu}\right)+\nonumber\\
&-&\frac{1}{p^{4}}\left(2(d-1)p_{\mu}p_{\nu}\right)\Bigg]\label{AAAk0}.
\end{eqnarray}
Thus, adding eqs.(\ref{AAAA}), (\ref{Acck0}) and (\ref{AAAk0}) we obtain
\begin{eqnarray}
i\Pi_{\mu\nu}^{(2)ab}(k)=2g^{2}N\delta^{ab}\int{\frac{d^{d}p}{(2\pi)^{d}}}\Bigg(\frac{(p.p)\delta_{\mu\nu}-2p_{\mu}p_{\nu}}{p^{2}(p-k)^{2}}\Bigg)\label{Sumk0}.
\end{eqnarray}
As can be seen, the dependence of the total self-energy on the gauge parameter $\alpha$ is eliminated by employing the HTL approximation. 

%

With this information, we can determine the thermal masses of the longitudinal and transverse sectors of the propagator of the Yang-Mills theory, given by \begin{eqnarray}
\Pi_{L}&=&\frac{K^{2}}{k^{2}}\Pi_{44}\label{regras1},\\
\Pi_{T}&=&\frac{1}{2}(\Pi_{\mu\mu}-\Pi_{L}).\label{regras2}
\end{eqnarray}
Therefore the gluon propagator at finite temperatures is,
\begin{eqnarray}
D_{\mu\nu}(K)=\frac{(P_{T})_{\mu\nu}}{K^{2}+\Pi_{T}}+\frac{(P_{L})_{\mu\nu}}{K^{2}+\Pi_{L}}+\alpha\frac{K_{\mu}K_{\nu}}{K^{4}},
\end{eqnarray}
with $(P_T)_{\mu\nu}$ and $(P_L)_{\mu\nu}$ being the transversal and longitudinal projectors, respectively, in the HTL approximation and where $K_{\mu}=(k_{4},\vec{k})=(-\omega_{n},\vec{k})$. 


The component $\mu=\nu$ of the total self-energy (\ref{Sumk0}) is,
\begin{eqnarray}
i\Pi^{(2)}_{\substack{\mu\mu \text{Total}}}(k) &=&4g^{2}N\int{\frac{d^{d}p}{(2\pi)^{d}}}\frac{1}{p^{2}}\nonumber\\
&=&4g^{2}NT\sum_{n}\int{\frac{d^{3}p}{(2\pi)^{3}}}\frac{1}{\omega_{n}^{2}+p^{2}}\nonumber\\
&=&\frac{-2}{2i\pi}\int{\frac{d^{3}p}{(2\pi)^{3}}}\int_{-i\infty+\delta}^{{i\infty+\delta}}{dp_{0}}\Bigg(\frac{4g^{2}Nn_{B}(p_{0})}{p_{0}^{2}+p^{2}}\Bigg)\nonumber\\
&=&\frac{4g^{2}N}{2\pi^{2}}\int_{0}^{\infty}{dp}\Bigg(\frac{p}{e^{\beta p}-1}\Bigg)\nonumber\\
&=&\frac{g^{2}NT^{2}}{3}\label{Sumk0mumu}.
\end{eqnarray}
where we considered that the dimension is $\delta_{\mu\mu}=d=4$, $n_{B}(p)=\frac{1}{e^{\beta p}-1}$, $\beta=\frac{1}{T}$ and $T$ is the temperature.

The component $(\mu,\nu)=(4,4)$ of the total self-energy (\ref{Sumk0}) is,
\begin{eqnarray}
i\Pi^{(2)}_{44 \mathrm{Total}}(k)&=&g^{2}N\int{\frac{d^{d}p}{(2\pi)^{d}}}\Bigg(\frac{-6}{(p-k)^{2}}+\frac{4(\vec{p}\cdot\vec{p})}{p^{2}(p-k)^{2}}\Bigg)\nonumber\\
&=&g^{2}N\int{\frac{d^{d}p}{(2\pi)^{d}}}\Bigg(\frac{-6}{p^{2}}+\frac{4(\vec{p}\cdot\vec{p})}{p^{2}(p-k)^{2}}\Bigg)\label{Sumk044}.
\end{eqnarray}
where we used that $\delta_{44}=-1$, $p_{4}^{2}=-\omega^{2}$ and $\omega^{2}=-p^{2}+\vec{p}\cdot\vec{p}$.
In order to solve these integrals we consider that,
\begin{eqnarray}
I&=&T\sum_{n}\int{\frac{d^{3}p}{(2\pi)^{3}}}\Bigg(\frac{\vec{p}\cdot\vec{p}}{(\omega_{n}^{2}+p^{2})((\omega_{n}-\omega_{l})^{2}+(p-k)^{2})}\Bigg)\nonumber\\
&=&\frac{-1}{8\pi^{2}}\int_{0}^{\infty}{dp}\Bigg[p^{2}\int_{-1}^{1}dy\Bigg(\frac{y}{y+\frac{i\omega_{l}}{k}}\frac{\partial n_{B}(p)}{\partial p}-\frac{1}{p}n_{B}(p)\Bigg)\Bigg]\nonumber\\
&=&\frac{T^{2}}{24}\Bigg(3-\frac{i\omega_{l}}{k}\log{\frac{\frac{i\omega_{l}}{k}+1}{\frac{i\omega_{l}}{k}-1}}\Bigg)
\end{eqnarray}
Then, eq.(\ref{Sumk044}) is rewritten as,
\begin{eqnarray}
i\Pi^{(2)}_{44 \text{Total}}(k)&=&\frac{-g^{2}NT^{2}}{2}+\frac{g^{2}NT^{2}}{6}\Bigg(3-\frac{i\omega_{l}}{k}\log{\frac{\frac{i\omega_{l}}{k}+1}{\frac{i\omega_{l}}{k}-1}}\Bigg)\nonumber\\
&=&\frac{g^{2}NT^{2}}{6}\Bigg(-\frac{i\omega_{l}}{k}\log{\frac{\frac{i\omega_{l}}{k}+1}{\frac{i\omega_{l}}{k}-1}}\Bigg)
\label{Sumk0441}.
\end{eqnarray}
Equations (\ref{Sumk0mumu}) and (\ref{Sumk0441}) show that the temperature dependence of the longitudinal and transversal sectors is provided by,
%
%
\begin{eqnarray}
\Pi_{L}&=&\frac{g^{2}NT^{2}}{3}\frac{K^{2}}{k^{2}}\Bigg(-\frac{i\omega_{l}}{2k}\log{\frac{\frac{i\omega_{l}}{k}+1}{\frac{i\omega_{l}}{k}-1}}\Bigg),\\
\Pi_{T}&=&\frac{g^{2}NT^{2}}{6}\Bigg[1-\frac{K^{2}}{k^{2}}\Bigg(-\frac{i\omega_{l}}{2k}\log{\frac{\frac{i\omega_{l}}{k}+1}{\frac{i\omega_{l}}{k}-1}}\Bigg)\Bigg].
\end{eqnarray}
Combining the Minkowski four-momentum $K_{\mu}=(\omega,\vec{k})$ and the analytic continuation $i\omega\rightarrow\omega+i\delta$, we obtain the propagator,
%
\begin{eqnarray}
D_{\mu\nu}(K)=-\frac{(P_{T})_{\mu\nu}}{K^{2}-\Pi_{T}}-\frac{(P_{L})_{\mu\nu}}{K^{2}-\Pi_{L}}-\alpha\frac{K_{\mu}K_{\nu}}{K^{4}},\label{prop finite t}
\end{eqnarray}
with,
\begin{eqnarray}
\Pi_{L}&=&\frac{g^{2}NT^{2}}{3}\frac{K^{2}}{k^{2}}\Bigg(1+i\pi\Theta\left(1-\frac{\omega}{k}\right)-\frac{\omega}{2k}\log{\frac{\frac{\omega}{k}+1}{\frac{\omega}{k}-1}}\Bigg),\\
\Pi_{T}&=&\frac{g^{2}NT^{2}}{6}\Bigg[1-\frac{K^{2}}{k^{2}}\Bigg(1+i\pi\Theta\left(1-\frac{\omega}{k}\right)-\frac{\omega}{2k}\log{\frac{\frac{\omega}{k}+1}{\frac{\omega}{k}-1}}\Bigg)\Bigg],\nonumber\\
\end{eqnarray}
and,
\begin{eqnarray}
(P_{T})_{\mu\nu}&=&\delta^{i}_{\mu}\delta^{j}_{\nu}\Bigg(\delta_{ij}-\frac{k_{i}k_{j}}{\vec{k}^{2}}\Bigg),\nonumber\\
(P_{L})_{\mu\nu}&=&\delta_{\mu\nu}-\frac{k_{\mu}k_{\nu}}{k^{2}}-(P_{T})_{\mu\nu}.
\end{eqnarray}
Using the series expansion $\frac{\omega}{2k}\log{\frac{\frac{\omega}{k}+1}{\frac{\omega}{k}-1}}=1+\frac{k^{2}}{3\omega^{2}}+O(\frac{1}{\omega^{3}})$ and having that for $(1-\frac{\omega}{k})<0\rightarrow\Theta(1-\frac{\omega}{k})=0$ we find that, after taking the limit $\frac{\omega}{k}\rightarrow\infty$,
\begin{eqnarray}
\Pi_{L}(\omega,k=0)&=&\Pi_{T}(\omega,k=0)=\frac{g^{2}NT^{2}}{9}=\frac{\omega_{D}^{2}}{3}\equiv\omega_{pl}^{2},\nonumber\\
D_{L,T}(\omega,k=0)&=&\frac{1}{\omega^{2}-\omega_{pl}^{2}}\label{result1},
\end{eqnarray}
where $\omega_{D}^{2}=\frac{g^{2}NT^{2}}{3}$ is the Debye screening mass, $\omega_{pl}$ is the plasma frequency and $D_{L,T}(K)\equiv(K^{2}-\Pi_{L,T}(K))^{-1}$ is the form factor of the longitudinal and transversal sectors of the gluon propagator at finite temperature. This indicates that the hot plasma's longitudinal and transverse non-static gluon propagators oscillate with a distinctive frequency of $\omega_{pl}=\frac{\omega_{D}}{\sqrt{3}}$, which we will refer to from here on out as simply the plasma mass, $M\equiv\omega_{pl}$.
%
%
%
In order to obtain the same result for the gluon propagator, we use the information in eq.(\ref{result1}) to construct an effective action for the description of the Yang-Mills theory at finite temperatures. To simplify things, we can simply include the term $\frac{1}{2}M^{2}A_{\mu}^{a}A_{\mu}^{a}$ into the action (\ref{actionLC}), so that,
%
\begin{eqnarray}
S&=&S_{YM}+S_{FP}^{LC}+\int{d^{4}x}\Bigg(\frac{1}{2}M^{2}A_{\mu}^{a}A_{\mu}^{a}\Bigg).\label{actionLCm}
\end{eqnarray}
From such action, it is easy to see that the gluon propagator will be,
\begin{eqnarray}
 \langle A_{\mu}^{a}(k)A_{\nu}^{b}(-k) \rangle=\Bigg[\frac{1}{k^{2}+M^{2}}\Bigg(\delta_{\mu\nu}-\frac{k_{\mu}k_{\nu}}{k^{2}}\Bigg)+\alpha\frac{k_{\mu}k_{\nu}}{k^{4}}\Bigg]\delta^{ab}.\nonumber\\\label{prop lin1}
\end{eqnarray}

\section{The Gribov Problem in the Landau gauge}\label{section4}
%
%
%
%
%
As previously observed in \ref{sec2}, in order to advance with the path integral quantization of a gauge theory, it is necessary to incorporate the gauge-fixing procedure formulated by Faddeev-Popov. This procedure is essential for circumventing the summation over an infinite number of equivalent gauge field configurations that arise from the gauge symmetry inherent in such theories. It is imperative to undertake this action, as failure to do so would result in the disruption of the route integral (\ref{z1}), leading to its divergence.  
Nevertheless, it is crucial to note that this technique is rigorously specified just within the context of studying the perturbative domain of the theory. Venturing into the non-perturbative regime, however, gives rise to certain challenges that indicate the fundamental nature of the problem is far from being fully understood.
The primary concern is around the non-invertibility of the Faddeev-Popov operator (\ref{FP operator}) when the amplitudes of the gauge fields $A_{\mu}^{a}$ attain large values. This phenomenon can be attributed to the appearance of the zero-modes of the Faddeev-Popov operator.
The comprehension of this matter can be facilitated through the examination of the eigenvalue equation of the Faddeev-Popov operator within the context of the Landau gauge, as deduced from, 
\begin{eqnarray}
M^{ab}\xi^{b}=-\partial_{\mu}D_{\mu}^{ab}\xi^{b}&=&\varepsilon(A)\xi^{b}\nonumber,\\
-\delta^{ab}\partial^{2}\xi^{b}+gf^{abc}\partial_{\mu}\left(A_{\mu}^{c}\xi^{b}\right)&=&\varepsilon(A)\xi^{b},\label{eigenvalue eq}
\end{eqnarray}
where $\varepsilon(A)$ are the eigenvalues of $M^{ab}$.
The term $-\delta^{ab}\partial^{2}$ in the equation mentioned exhibits positive eigenvalues, except for the trivial solution. This implies that for $g=0$, the Faddeev-Popov operator (\ref{FP operator}) must be invertible. However, it should be noted that the second term, which is expressed as $gf^{abc}(\partial_{\mu}A_{\mu}^{c})+gf^{abc}A_{\mu}^{c}\partial_{\mu}$, exhibits negative eigenvalues. Hence, when the values of $gA_{\mu}$ are sufficiently large, it is possible to observe negative eigenvalues for the Faddeev-Popov operator. This implies that, at a certain stage, the zero-modes of the operator have been inevitably attained. This implies that, in the regime of high values of the coupling constant $g$ when perturbation theory is not applicable, an alternative methodology may be employed to address this issue.



Significant advancements were achieved by Zwanziger in \cite{dell1991every}, wherein the Faddeev-Popov quantization procedure was enhanced by imposing limitations on the integration domain of the path integral (\ref{z1}). These limitations were applied to a specific region within the gauge field configuration space, ensuring that all $A_{\mu}^{a}$ fields satisfy the gauge-fixing conditions and that the Faddeev-Popov operator possesses invertibility.
This region is the so-called Gribov region which presents important features like:
\begin{itemize}
    \item It is bounded in all directions (in the Landau gauge);
    \item It is convex;
    \item All the gauge orbits cross the Gribov region at least once.
\end{itemize}
The third one ensures that there is no loss of physical information throughout the process of limiting the path integral to the Gribov region. This phenomenon arises from the fact that every gauge field configuration existing beyond the Gribov region is inherently a gauge copy of a configuration contained within it and, for this reason, are called as Gribov copies. 



For the Landau gauge the Gribov region $\Omega$ will be defined as,
\begin{eqnarray}
\Omega=\lbrace A_{\mu}^{a}|\partial_{\mu}A_{\mu}^{a}=0,M^{ab}>0\rbrace.
\end{eqnarray}

Then, in order to restrict the path integral (\ref{z1}) to the region $\Omega$, we will implement the condition that the ghost propagator does not develop poles other than at $k=0$ which will be the place where occurs the first zero-mode of the Faddeev-Popov operator $M^{ab}$. This is also called as the 'no-pole condition' and the first zero-mode is known as the Gribov horizon. Thus we have that,
%
\begin{eqnarray}
\langle \bar{c}^{a}(k)c^{b}(-k)\rangle=\langle k|(M^{-1}(A))^{ab}|k\rangle,
\end{eqnarray}
which, up to one loop is given by,
\begin{eqnarray}
\langle \bar{c}^{a}(k)c^{b}(-k)\rangle\approx\frac{1}{k^{2}}\frac{1}{1-\rho(k,A)},
\end{eqnarray}
where,
\begin{eqnarray}
\rho(k,A)=\frac{N}{N^{2}-1}\frac{g^{2}}{V}\frac{k_{\mu}k_{\nu}}{k^{2}}\sum_{q}\frac{A_{\mu}^{a}(q)A_{\nu}^{a}(-q)}{(k-q)^{2}}.
\end{eqnarray}
We can see that at $\rho(k,A)=1$ we find a second pole for the ghost propagator, that is associated to a second zero-mode of the Faddeev-Popov operator. This essentially means that we are reaching gauge field configurations that lie outside of the Gribov region $\Omega$. 
In other words, we need to limit the integration domain of the path integral to the gauge fied configurations for which $\rho(k,A)<1$ in order to mantain it whithin the Gribov region.
Moreover, as $\rho(k,A)$ is a monotonically decreasing function, we have that such a condition can be restricted to,
\begin{eqnarray}
\rho(0,A)<1.\label{no pole cond}
\end{eqnarray}
Finally, in order to apply this condition to the path integral (\ref{z1}) we will we will make use of a Heaviside function $\Theta(1-\rho(0,A))$ resulting in,
%
\begin{eqnarray}
    Z&=&\mathcal{N}\int[DA][D\bar{c}][Dc]\Theta(1-\rho(0,A))e^{-S},\nonumber\\
    &=&\mathcal{N}\int_{-i\infty+\epsilon}^{{i\infty+\epsilon}}\frac{d\zeta}{2\pi i\zeta}\int{[DA]}[D\bar{c}][Dc]e^{-S+\zeta(1-\rho(0,A))},\nonumber\\\label{z2}
\end{eqnarray}
with $S$ given by eq.(\ref{actionLC}) and $\zeta$ is a function that we used in order to rewrite the Heaviside theta in an integral form. 

The quadratic part over the fields of eq.(\ref{z2}) will be given by,
\begin{eqnarray}
Z^{quadr}&=&\mathcal{N}\int\frac{d\zeta}{2\pi i}e^{f(\zeta)},\nonumber\\
f(\zeta)&=&\zeta-ln\zeta-\frac{3}{2}(N^{2}-1)\sum_{q}ln\left(q^{2}+\frac{Ng^{2}\zeta}{2Vq^{2}(N^{2}-1)}\right)\nonumber.\\
\end{eqnarray}
In its saddle point we have that,
\begin{eqnarray}
Z^{quadr}&=&e^{f(\zeta_{0})}.
\end{eqnarray}
Here, we have that $\zeta_{0}$ is the saddle point of the function $f(\zeta)$ or, in other words,
\begin{eqnarray}
f'(\zeta_{0})=0\rightarrow1=\frac{1}{\zeta_{0}}+\frac{3}{4V}Ng^{2}\sum_{q}\frac{1}{q^{4}+\frac{\zeta_{0}Ng^{2}}{2V(N^{2}-1)}}.
\end{eqnarray}
Now, defining the so-called 'Gribov parameter' $\gamma^{4}=\frac{\zeta_{0}}{4V(N^{2}-1)}$ we have that,
\begin{eqnarray}
1=\frac{1}{4V(N^{2}-1)\gamma^{4}}+\frac{3}{4V}Ng^{2}\sum_{q}\frac{1}{q^{4}+2Ng^{2}\gamma^{4}}.\label{pre gap}
\end{eqnarray}
Then by applying the thermodynamic limit $V\rightarrow\infty$ we obtain,
\begin{eqnarray}
1=\frac{3}{4}Ng^{2}\int\frac{d^{4}q}{(2\pi)^{4}}\frac{1}{q^{4}+2Ng^{2}\gamma^{4}},\label{gap gribov}
\end{eqnarray}
which is also referred to as the 'gap equation'. This equation allows us to calculate the Gribov parameter as a function of the coupling constant $g$ and the ultraviolet cut-off $\Lambda$. 
%

Now, by calculating the gluon propagator from the path integral (\ref{z2}) we obtain,
%
\begin{eqnarray}
\langle A_{\mu}^{a}(k)A_{\nu}^{b}(-k)\rangle=\frac{k^{2}}{k^{4}+2Ng^{2}\gamma^{4}}\Bigg(g_{\mu\nu}-\frac{k_{\mu}k_{\nu}}{k^{2}}\Bigg)\delta^{ab}. \label{prop gluon gribov}
\end{eqnarray}
By analysing the propagator above, it can be observed that its structure indicates the presence of a complex pole, implying the existence of non-physical degrees of freedom, since the mass of the gluons will fall beyond the physical spectrum. This property holds significant importance as it enables us to understand it as the propagator of confined gluons, given that confined particles must necessarily have non-physical masses since they cannot be directly measured. 



Before moving on to the next part, it must be said that the restriction done to the path integral on eq. (\ref{z2}) was done by imposing the no-pole condition up to the first order of the ghost two-point function. But an all-order prescription was first made at \cite{zwanziger1989local}, where it was imposed that the smallest eigenvalue of the Faddeev-Popov operator could only be positive. This means that the Gribov region will be set by the following condition in the Landau gauge,
%
%
\begin{eqnarray}
\varepsilon_{min}(A)\geq0.
\end{eqnarray}
This implies that,
\begin{eqnarray}
tr(M^{ab}(A))=Vd(N^{2}-1)-H_{L}[A]>0,
\end{eqnarray}
where $H_{L}[A]$ is the so-called 'horizon function'. This non-local function is what in fact will restrict our path integral to the Gribov region at all orders in the series expansion. In its explicit form it is given by, 
\begin{eqnarray}
H_{L}[A]&=&\int{d^{4}x d^{4}y}\Bigg(g^{2}f^{abc}A_{\mu}^{b}(x)[M^{-1}(A)]^{ad}(x,y)\times\nonumber\\
&\times&f^{dec}A_{\mu}^{e}(y)\Bigg).
\end{eqnarray}
Therefore the total action of the Yang-Mills theory in the Landau gauge taking into account the infinitesimal Gribov ambiguities becomes, 
\begin{eqnarray}
S_{GZ}&=&S_{YM}+S_{FP}^{L}+\int{d^{4}x}\Bigg(\gamma^{4}H_{L}[A]+\nonumber\\
&+&4\gamma^{4}V(N^{2}-1)\Bigg),\nonumber\\
\end{eqnarray}
that is also known as the Gribov-Zwanziger action.

In order to localize the horizon function we will make use of two pair of bosonic and fermionic auxiliary fields, respectively $(\bar{\phi},\phi)$ and $(\bar{\omega},\omega)$. We will do so by means of the following identity,
\begin{eqnarray}
e^{-\gamma^{4}H_{LC}[A]}&=&\int{[D\bar{\phi}][D\phi][D\bar{\omega}][D\omega]}\times\nonumber\\
&\times&exp\Bigg[\int{d^{4}x}d^{4}y\Bigg(\bar{\phi}_{\mu}^{ac}(x)M^{ab}(x,y)\phi_{\mu}^{bc}(y)+\nonumber\\
&-&\bar{\omega}_{\mu}^{ac}(x)M^{ab}(x,y)\omega_{\mu}^{bc}(y)\Bigg)+\nonumber\\
&-&g\gamma^{2}\int{d^{4}x}f^{abc}A_{\mu}^{a}(x)\left(\bar{\phi}_{\mu}^{bc}(x)+\phi_{\mu}^{bc}(x)\right)\Bigg].\nonumber\\
\end{eqnarray}
Thus, the total Gribov-Zwanziger action becomes, 
\begin{eqnarray}
S_{GZ}&=&S_{YM}+S_{FP}^{L}+\nonumber\\
&-&\int{d^{4}x}\left(\bar{\phi}_{\mu}^{ac}M^{ab}(A)\phi_{\mu}^{bc}-\bar{\omega}_{\mu}^{ac}M^{ab}(A)\omega_{\mu}^{bc}\right)+\nonumber\\
&+&\int{d^{4}x}\left[g\gamma^{2}f^{abc}A_{\mu}^{a}\left(\bar{\phi}_{\mu}^{bc}+\phi_{\mu}^{bc}\right)+4\gamma^{4}V(N^{2}-1)\right].\nonumber\\\label{GZ nao Brst}
\end{eqnarray}
Despite being local, the action (\ref{GZ nao Brst}) breaks the BRST symmetry in the non-perturbative regime, as can be seen on \cite{capri2009aspectos,sorella2018simetria,capri2016local}. Then in order to reestablish it, one needs to make use of a non-local, transverse and gauge invariant field $A_{\mu}^{h,a}$ given by,
%
%
\begin{eqnarray}
A_{\mu}^{h}&=&\Bigg(\delta_{\mu\nu}-\frac{k_{\mu}k_{\nu}}{\partial^{2}}\Bigg)\Bigg(A_{\nu}-ig\Bigg[\frac{1}{\partial^{2}}\partial A,A_{\nu}\Bigg]+\nonumber\\
&+&\frac{ig}{2}\Bigg[\frac{1}{\partial^{2}}\partial A,\partial_{\nu}\frac{1}{\partial^{2}}\partial A\Bigg]+O(A^{3})\Bigg).\label{Ah}
\end{eqnarray}
Basically, what we do is to modify the horizon function changing every field $A_{\mu}^{a}$ by $A_{\mu}^{h,a}$, in such a way that the action becomes,
\begin{eqnarray}
S_{GZ}&=&S_{YM}+S_{FP}^{L}+\nonumber\\
&-&\int{d^{4}x}\left(\bar{\phi}_{\mu}^{ac}M^{ab}(A^{h})\phi_{\mu}^{bc}-\bar{\omega}_{\mu}^{ac}M^{ab}(A^{h})\omega_{\mu}^{bc}\right)+\nonumber\\
&+&\int{d^{4}x}\left[g\gamma^{2}f^{abc}A_{\mu}^{h,a}\left(\bar{\phi}_{\mu}^{bc}+\phi_{\mu}^{bc}\right)+4\gamma^{4}V(N^{2}-1)\right],\nonumber\\\label{RGZ Brst}
\end{eqnarray}
with,
\begin{eqnarray}
M^{ab}(A^{h})&=&-\delta^{ab}\partial^{2}+gf^{abc}A_{\mu}^{h,a}\partial_{\mu}.\label{Mab h}
\end{eqnarray}
As this new action (\ref{RGZ Brst}) is non-local, we need to rewrite the field $A_{\mu}^{h}$ in a local fashion using a stuekelberg field $\xi^{a}$ as in \cite{capri2016local} in such a way that,
\begin{eqnarray}
    A_{\mu}^{h}= h^{\dag}A_{\mu}h+\frac{i}{g}h^{\dag}\partial_{\mu}h,\label{hstuekelberg}
\end{eqnarray}
where,
\begin{eqnarray}
    h=e^{ig\xi^{a}T^{a}}.
\end{eqnarray}
Then we implement the transversality condition of the $A_{\mu}^{h,a}$ field,
\begin{eqnarray}
\partial_{\mu}A_{\mu}^{h,a}=0,
\end{eqnarray}
in such a way that its local version becomes,
\begin{eqnarray}
S_{GZ}&=&S_{YM}+S_{FP}^{L}+\nonumber\\
&-&\int{d^{4}x}\left(\bar{\phi}_{\mu}^{ac}M^{ab}(A^{h})\phi_{\mu}^{bc}-\bar{\omega}_{\mu}^{ac}M^{ab}(A^{h})\omega_{\mu}^{bc}\right)+\nonumber\\
&+&\int{d^{4}x}\left[g\gamma^{2}f^{abc}A_{\mu}^{h,a}\left(\bar{\phi}_{\mu}^{bc}+\phi_{\mu}^{bc}\right)+4\gamma^{4}V(N^{2}-1)\right]+\nonumber\\
&+&\int{d^{4}x}\left[\tau^{a}\partial_{\mu}A_{\mu}^{h,a}-\bar{\eta}^{a}M^{ab}(A^{h})\eta^{b}\right],
\label{RGZ Brst1}
\end{eqnarray}
where $\tau^{a}$ is a Lagrange multiplier to ensure the transversality condition and the fields $(\bar{\eta}^{a},\eta^{a})$ are ghost-like fields. 
As can be seen on \cite{capri2015exact,capri2016local,capri2017renormalizability} the action (\ref{RGZ Brst1}) after all these steps is finally BRST invariant.

\section{Extension to the linear covariant gauges}\label{extensionlc}
In order to extend the Gribov-Zwanziger action in Landau gauge to the linear covariant gauges we need to note first the differences between them. One of them is the fact that the Faddeev-Popov operator is not Hermitian except at the Landau limit $\alpha\rightarrow0$. This is important since this feature complicates the Gribov-Zwanziger analysis for eliminating infinitesimal Gribov copies from the path integral measure

despite being a special case for $\alpha=0$ of the gauge condition $\partial_{\mu}A_{\mu}^{a}=-i\alpha b^{a}$, the Landau gauge has the property of having an Hermitian Faddeev-Popov operator $M^{ab}$.
But thanks to the foundational work of \cite{capri2015exact,capri2017renormalizability}, which provided the basis for formulating a BRST-invariant GZ action in the Landau gauge, we can now address this problem within the context of linear covariant gauges as well. By repeating the steps of \cite{capri2015exact,capri2017renormalizability} we obtain the following local and BRST-invariant action for Yang-Mills theory in the linear covariant gauges within the Gribov-Zwanziger framework,
\begin{eqnarray}
S_{GZ}&=&S_{YM}+S_{FP}^{LC}+\nonumber\\
&-&\int{d^{4}x}\left(\bar{\phi}_{\mu}^{ac}M^{ab}(A^{h})\phi_{\mu}^{bc}-\bar{\omega}_{\mu}^{ac}M^{ab}(A^{h})\omega_{\mu}^{bc}\right)+\nonumber\\
&+&\int{d^{4}x}\left[g\gamma^{2}f^{abc}A_{\mu}^{h,a}\left(\bar{\phi}_{\mu}^{bc}+\phi_{\mu}^{bc}\right)+4\gamma^{4}V(N^{2}-1)\right]+\nonumber\\
&+&\int{d^{4}x}\left[\tau^{a}\partial_{\mu}A_{\mu}^{h,a}-\bar{\eta}^{a}M^{ab}(A^{h})\eta^{b}\right],
\label{RGZ lc1}
\end{eqnarray}
with $S_{FP}^{LC}$ being given by eq. (\ref{fplc1}).
Here, the path integral is being restricted to the region,
\begin{eqnarray}
\Omega=\lbrace A_{\mu}^{a}|\partial_{\mu}A_{\mu}^{a}=-i\alpha b^{a},M^{ab}(A^{h})>0\rbrace,
\end{eqnarray}
where $M^{ab}(A^{h})$ is given on eq. (\ref{Mab h}).
Finally, from this action we can calculate the gluon propagator,
\begin{eqnarray}
\langle A_{\mu}^{a}(k)A_{\nu}^{b}(-k)\rangle&=&\Bigg[\frac{k^{2}}{k^{4}+2Ng^{2}\gamma^{4}}\Bigg(\delta_{\mu\nu}-\frac{k_{\mu}k_{\nu}}{k^{2}}\Bigg)+\nonumber\\
&+&\alpha\frac{k_{\mu}k_{\nu}}{k^{4}}\Bigg]\delta^{ab}, \label{prop gluon gribov lc}
\end{eqnarray}
which reobtains the Landau one (\ref{prop gluon gribov}) after taking the limit $\alpha\rightarrow0$.



\section{Refined Gribov-Zwanziger in Linear covariant gauges}\label{section6}

Localization of the Gribov-Zwanziger action is achieved by introducing pairs of bosonic and fermionic auxiliary fields, as was demonstrated on section \ref{extensionlc}. From this approach we obtained the local and BRST-invariant action (\ref{RGZ lc1}) and we calculated the gluon propagator (\ref{prop gluon gribov lc}). 
by analysing it we can observe that in the limit $k\rightarrow0$ such propagator manifest a scaling behaviour going to zero in this limit. 

When compared with lattice result \cite{ cucchieri2008constraints, cucchieri2008constraints1, maas2009more} it was seen that this scaling behaviour of the gluon propagator obtained from the GZ action did not agreed with such previous works. Therefore, in order to solve this problem a complementary approach was developed. The idea was to build an effective BRST-invariant action that could capture the effects due to non-trivial condensates that arise due to instabilities of the GZ action in the infrared regime \cite{dudal2008new,dudal2008refinement}. 
\begin{eqnarray}
S_{O}=\int{d^{4}x}\left(JO-\rho g^{2}J\right),
\end{eqnarray}
where $O$ is a $2$-dimensional operator, $J$ is the source, and $\rho$  is a dimensionless parameter needed in order to grant the renormalizability of the theory since it is responsible for removing any potential divergencies proportional to $g^{2}J$. 
Given that the source $J$ possesses a dimension of mass squared, we can express $J$ as $M^{2}$. Here, $M^{2}$ represents the mass of the condensate $\langle O\rangle$, which influences the poles of the propagator, thereby altering its physical spectrum. The action derived from this process is referred to as the Refined Gribov-Zwanziger action (RGZ). In linear covariant gauges, the RGZ action is defined as follows,
\begin{eqnarray}
S_{RGZ}=S_{GZ}+S_{m}+S_{\mu},\label{S RGZ}
\end{eqnarray}
where,
\begin{eqnarray}
S_{m}&=&\int{d^{4}x}\Bigg(\frac{1}{2}m^{2}A_{\mu}^{h,a}A_{\mu}^{h,a}\Bigg),\\
S_{\mu}&=&\int{d^{4}x}\Bigg(-\mu^{2}(\bar{\phi}_{\mu}^{ab}\phi_{\mu}^{ab}-\bar{\omega}_{\mu}^{ab}\omega_{\mu}^{ab})\Bigg).
\end{eqnarray}
are respectively the actions that give the contribution of the condensation of the $2$-dimensional operators $A_{\mu}^{h,a}A_{\mu}^{h,a}$ and $\bar{\phi}^{ab}_{\mu}\phi^{ab}_{\mu}-\bar{\omega}^{ab}_{\mu}\omega^{ab}_{\mu}$.
From equation (\ref{S RGZ}), we deduce that the gluon propagator, when expressed in linear covariant gauges, takes the form,
\begin{eqnarray}
 \langle A_{\mu}^{a}(k)A_{\nu}^{b}(-k) \rangle&=&\Bigg[\frac{k^{2}+\mu^{2}}{(k^{2}+m^{2})(k^{2}+\mu^{2})+2Ng^{2}\gamma^{4}}\times\nonumber\\
 &\times&\Bigg(\delta_{\mu\nu}-\frac{k_{\mu}k_{\nu}}{k^{2}}\Bigg)+\frac{\alpha k_{\mu}k_{\nu}}{k^{4}}\Bigg]\delta^{ab}.\label{prop RGZ lin}
\end{eqnarray}
It's crucial to note that the condensate masses $m^{2}$ and $\mu^{2}$ are each defined by their respective gap equations, as outlined in \cite{ferreira2021infrared}. However, this work does not aim to solve these equations. Instead, our focus is on a qualitative analysis of the gluon propagator's behavior at finite temperatures, which will be discussed in subsequent sections. For the Gribov parameter $\gamma^{4}$, within the RGZ framework, it is determined by the following gap equation,
%
%
\begin{eqnarray}
1=\frac{3}{4}Ng^{2}\int\frac{d^{4}q}{(2\pi)^{4}}\frac{1}{\left(q^{2}+m^{2}\right)\left(q^{2}+\mu^{2}\right)+2Ng^{2}\gamma^{4}},
\end{eqnarray}
which was derived following the same steps that we took to derive eq. (\ref{gap gribov}).

\section{The effective thermal massive model}\label{section7}
\subsection{Effective formulation $1$}
As we concluded in section \ref{sec 3}, the study of Yang-Mills theory at finite temperatures necessitates the construction of an effective action. This action allows us to derive the gluon propagator (\ref{prop finite t}). The addition of a term such as $\frac{1}{2}M^{2}A_{\mu}^{a}A_{\mu}^{a}$ to the theory's action is all that's required. Consequently, for the RGZ scenario, our finite temperature effective action is as follows,
\begin{eqnarray}
S=S_{RGZ}+S_{M},\label{Slinearcov}
\end{eqnarray}
where,
\begin{eqnarray}
S_{M}=\int{d^{4}x}\Bigg(\frac{1}{2}M^{2}A_{\mu}^{a}A_{\mu}^{a}\Bigg).
\end{eqnarray}
is the effective term that adds the thermal mass contribution into the theory and $M^{2}=\frac{g^{2}NT^{2}}{9}$ is the plasma mass.

So the quadratic part of eq.(\ref{Slinearcov}) will be,
\begin{eqnarray}
S^{quadr}&=&\int{d^{4}x}\Bigg\{\frac{1}{2}A_{\mu}^{a}\Bigg[k^{2}\delta_{\mu\nu}-\left(1-\frac{1}{\alpha}\right)k_{\mu}k_{\nu}+\nonumber\\
&+&M^{2}\delta_{\mu\nu}+\left(\delta_{\mu\nu}-\frac{k_{\mu}k_{\nu}}{k^{2}}\right)\left(m^{2}+\frac{2Ng^{2}\gamma^{4}}{k^{2}+\mu^{2}}\right)\Bigg]\delta^{ab}A_{\nu}^{b}\Bigg\},\nonumber\\\label{SquadrLC}
\end{eqnarray}
from which we obtain the that the gluon propagator is given by,
\begin{widetext}
\begin{eqnarray}
 \langle A_{\mu}^{a}(k)A_{\nu}^{b}(-k) \rangle&=&\Bigg[\frac{k^{2}+\mu^{2}}{(k^{2}+m^{2}+M^{2})(k^{2}+\mu^{2})+2Ng^{2}\gamma^{4}}\times\nonumber\\
 &\times&\Bigg(\delta_{\mu\nu}-\frac{k_{\mu}k_{\nu}}{\alpha M^{2}+k^{2}}\Bigg)+\frac{\alpha k_{\mu}k_{\nu}}{k^{2}\left(\alpha M^{2}+k^{2}\right)}+\nonumber\\
 &-&\frac{\alpha k_{\mu}k_{\nu}M^{2}\left(k^{2}+\mu^{2}\right)}{k^{2}\left(\alpha M^{2}+k^{2}\right)\left[\left(k^{2}+M^{2}+m^{2}\right)\left(k^{2}+\mu^{2}\right)+2Ng^{2}\gamma^{4}\right]}\Bigg]\delta^{ab}.\label{prop lin}
\end{eqnarray}
\end{widetext}
It's straightforward to observe that by setting $\alpha\rightarrow0$, we revert to the Landau result. Similarly, by setting $M\rightarrow0$, we obtain the result at $0$-temperature.

We can also see that the thermal mass $M^{2}$ impacts not only the transverse sector of the propagator, but also the longitudinal one. This is because the effective term $S_{M}$ in the action is expressed in terms of the fields $A_{\mu}^{a}$, rather than the non-local transverse fields $A_{\mu}^{h,a}$.
Hence, by applying the limit $\gamma\rightarrow0$ to the propagator (\ref{prop lin}), it becomes clear that it does not retrieve the same propagator as in eq. (\ref{prop finite t}). This implies that the current method may not be the best fit for describing the gluon infrared behaviour at finite temperatures. Consequently, we will explore a slightly different approach in the following section.

\subsection{Effective Formulation $2$}
In contrast to the previous section, for the finite temperature study of the YM theory, we will incorporate an effective term $\frac{1}{2}M^{2}A_{\mu}^{h,a}A_{\mu}^{h,a}$ into the RGZ action. The field $A_{\mu}^{h,a}$ here is non-local, transverse, and gauge invariant, expressed as an infinite non-local series in eq. (\ref{Ah}). This strategy has the benefit of preserving the gauge invariance of the theory.

Similar to the previous instance, the total action of this effective model will be,
\begin{eqnarray}
S=S_{RGZ}+S_{M},\label{Slinearcov 2}
\end{eqnarray}
but with,
\begin{eqnarray}
S_{M}=\int{d^{4}x}\Bigg(\frac{1}{2}M^{2}A_{\mu}^{h,a}A_{\mu}^{h,a}\Bigg).
\end{eqnarray}
being the effective term that introduces the thermal mass contribution into the theory. As before, $M^{2}=\frac{g^{2}NT^{2}}{9}$ is the plasma mass.

Therefore, the quadratic part of eq.(\ref{Slinearcov 2}) will be,
\begin{eqnarray}
S^{quadr}&=&\int{d^{4}x}\Bigg\{\frac{1}{2}A_{\mu}^{a}\Bigg[k^{2}\delta_{\mu\nu}-\left(1-\frac{1}{\alpha}\right)k_{\mu}k_{\nu}+\nonumber\\
&+&\left(\delta_{\mu\nu}-\frac{k_{\mu}k_{\nu}}{k^{2}}\right)\left(M^{2}+m^{2}+\frac{2Ng^{2}\gamma^{4}}{k^{2}+\mu^{2}}\right)\Bigg]\delta^{ab}A_{\nu}^{b}\Bigg\},\nonumber\\\label{SquadrLC 2}
\end{eqnarray}
from which we deduce that the gluon propagator is given by,
\begin{eqnarray}
 \langle A_{\mu}^{a}(k)A_{\nu}^{b}(-k) \rangle&=&\Bigg[\frac{k^{2}+\mu^{2}}{(k^{2}+m^{2}+M^{2})(k^{2}+\mu^{2})+2Ng^{2}\gamma^{4}}\times\nonumber\\
 &\times&\Bigg(\delta_{\mu\nu}-\frac{k_{\mu}k_{\nu}}{k^{2}}\Bigg)+\alpha\frac{k_{\mu}k_{\nu}}{k^{4}}\Bigg]\delta^{ab}.\label{prop lin 2}
\end{eqnarray}
By setting $\alpha\rightarrow0$, it is observed that the propagator (\ref{prop lin 2}) reverts to the Landau result. Similarly, by setting $M\rightarrow0$, we retrieve the $0$-temperature result.

Unlike eq. (\ref{prop lin}), the propagator (\ref{prop lin 2}) has its longitudinal sector unaffected by the thermal mass $M$, with only the transverse one being affected. This is attributed to the transversality property of the field $A_{\mu}^{h,a}$. Consequently, by taking the limit $\gamma\rightarrow0$, we find that the propagator (\ref{prop lin}) reacquires the propagator of eq. (\ref{prop finite t}). This suggests that this alternative effective formulation is a promising candidate for depicting the infrared behaviour of gluons at finite temperatures and, for this reason, it will be the model under consideration in our subsequent sections.

%

\section{Analysis of the poles of the propagator}\label{section8}
With the gluon propagator (\ref{prop lin 2}) in hand, our next step is to dissect its physical spectrum. This will allow us to distinguish between the confining and deconfining regimes of the theory. 

Our approach involves examining the poles of the propagator, which reveal the masses of the gluonic degrees of freedom. Depending on these mass values, we can determine whether they correspond to physical or unphysical degrees of freedom. Essentially, a real and positive pole indicates a physical degree of freedom, while a negative or complex pole signifies an unphysical degree of freedom.

In the end, we attribute the physical degrees of freedom to the deconfined regime, and the unphysical ones to the confined regime.


This phenomenon arises from the presence of imaginary poles, which result in a violation of the positivity condition in the spectral density function of the Källén–Lehmann representation. This violation is indicative of the confinement of the associated degrees of freedom, as asymptotically free states are not expected to be observed in such a regime, as evidenced by \cite{peskin1995introduction,alkofer2001infrared}. 


Considering this, we can rewrite the gluon propagator (\ref{prop lin 2}) as follows,
%
\begin{eqnarray}
 \langle A_{\mu}^{a}(k)A_{\nu}^{b}(-k) \rangle=-\frac{\frac{\mu ^2+u_{1}}{k^2-u_{1}}-\frac{\mu ^2+u_{2}}{k^2-u_{2}}}{\sqrt{\left(-\mu ^2+m^2+M^2\right)^2-8Ng_{0}^{2}\gamma^{4}}},
\end{eqnarray}
where,
\begin{eqnarray}
u_{1}&=&-\frac{1}{2} \Bigg(\sqrt{\left(-\mu ^2+m^2+M^2\right)^2-8 Ng_{0}^{2}\gamma^{4}}+\mu^2+m^2+M^2\Bigg),\nonumber\\
u_{2}&=&\frac{1}{2} \Bigg(\sqrt{\left(-\mu ^2+m^2+M^2\right)^2-8 Ng_{0}^{2}\gamma^{4}}-\mu ^2-m^2-M^2\Bigg).\nonumber\\\label{polos}
\end{eqnarray}
We can see that,
\begin{itemize}
    \item For $\left(-\mu ^2+m^2+M^2\right)^2<8 Ng_{0}^{2}\gamma^{4}$, or $-2 \sqrt{2N}g_{0}\gamma^{2}+\mu ^2-m^2<M^{2}< 2 \sqrt{2N}g_{0}\gamma^{2}+\mu ^2-m^2$, both poles turn out to be complex. This implies a confinement regime for all degrees of freedom.
    \item For $\left(-\mu ^2+m^2+M^2\right)^2\geq8 Ng_{0}^{2}\gamma^{4}$, or $2 \sqrt{2N}g_{0}\gamma^{2}+\mu ^2-m^2<M^{2}<-2 \sqrt{2N}g_{0}\gamma^{2}+\mu ^2-m^2$, we observe two real poles, but only one is positive. This indicates the presence of a single deconfined degree of freedom, suggesting a regime of partial confinement-deconfinement.
    \item In the absence of a solution for the gap equation, the only viable option for the Gribov mass parameter is $\gamma=0$, which results in a free gluon propagator (deconfined phase).
\end{itemize}


Given that $M^{2}=\frac{g^{2}NT^{2}}{9}$, the above relations provide us with the critical temperatures at which each of the three regimes takes place.

When we compare our findings with Canfora's paper \cite{canfora2014gribov}, we discover that our results align with his in the limit where $m,\mu\rightarrow0$.

It's worth noting that if we compute the longitudinal and transverse thermal masses considering both the Gribov parameter and the condensate masses, we would encounter a more intricate pole structure and potentially additional confining and deconfining regimes. However, this topic will be explored in future discussions.




\section{The RGZ Gap equation in Linear covariant gauges}\label{section9}
An alternative perspective for analyzing the phase transitions of the theory is by scrutinizing the RGZ gap-equation at finite temperatures. This is significant because it enables us to compute the relationship between the Gribov parameter $\gamma^{4}$ and the temperature $T$. If we adopt a consistent procedure, the solutions should converge to the zero-temperature solution in the limit as $T\rightarrow0$. In essence, we should procure solutions that depict confined gluons in the limit as $T\rightarrow0$, and for sufficiently high temperatures, we should not discern any solution for the gap-equation, signifying a regime of freely propagating gluons.



In order to obtain the gap equation, we will follow again the steps of \cite{ferreira2021infrared}, that we used to derive the previous gap equation (\ref{gap gribov}), where we obtain that,
\begin{eqnarray}
1&=&\frac{3}{4V}Ng^{2}\sum_{q}\frac{1}{\left(q^{2}+\mu^{2}\right)\left(q^{2}+M^{2}+m^{2}+\frac{\gamma^{4}}{\left(q^{2}+\mu^{2}\right)}\right)}+\nonumber\\
&+&\frac{1}{4V(N^{2}-1)\gamma^{4}},
\end{eqnarray}
which can be rewritten as,
\begin{eqnarray}
1&=&\frac{3Ng^{2}\lambda}{16\pi^{3}}\sum_{n}\int_{0}^{1}dR\Bigg\{R^{2}\Bigg[(R^{2}+\theta_{n}^{2})^{2}+\nonumber\\
&+&\left(m^{2}+\mu^{2}+\frac{Ng^{2}\lambda^{2}}{36\pi^{2}}\right)(R^{2}+\theta_{n}^{2})+\nonumber\\
&+&\mu^{2}\left(m^{2}+\frac{Ng^{2}\lambda^{2}}{36\pi^{2}}\right)+\Gamma^{4}\Bigg]^{-1}\Bigg\},\label{gap no landau rgz}
\end{eqnarray}
where we used that the thermal mass is given by
\begin{eqnarray}
M^{2}=\frac{Ng^{2}T^{2}}{9}=\frac{Ng^{2}\lambda^{2}}{36\pi^{2}},
\end{eqnarray}
and we used the following parametrizations,
\begin{eqnarray}
R&=&\frac{r}{\Lambda},\nonumber\\
\lambda&=&\frac{2\pi T}{\Lambda},\nonumber\\
\theta_{n}&=&\frac{\omega_{n}}{\Lambda}=n\lambda ,\nonumber\\
\Gamma&=&\frac{\left(2Ng^{2}\right)^{1/4}\gamma}{\Lambda}.
\end{eqnarray}
It is evident that the gap equation remains identical to the Landau one. This implies that the gap equation is independent of the gauge parameter $\alpha$.


Now we can rewrite eq. (\ref{gap no landau rgz}) in the following way,
\begin{eqnarray}
F(\lambda,\Gamma,m,\mu)=1,
\end{eqnarray}
where,
\begin{eqnarray}
F(\lambda,\Gamma,m,\mu)=\frac{3Ng^{2}\lambda}{16\pi^{3}}\int_{0}^{1}{dR}\Bigg(R^{2}S(R,\lambda,\Gamma,m,\mu)\Bigg)=1,\nonumber\\\label{gap equation RGZ}
\end{eqnarray}
From this, we derive the Gribov parameter as a function of $\lambda, m, \mu$, and $g$.



\section{Analysis of the regimes of the theory}\label{section10}

\begin{figure}[t!]
\centering
\includegraphics[width=0.5\textwidth]{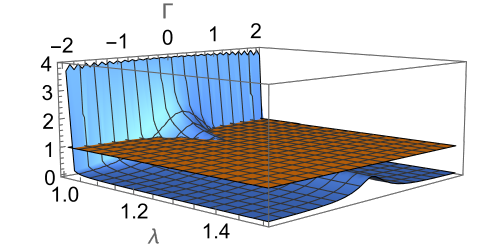}
\caption{\label{fig:1}Plot of the surface $F$ for
different values of $\lambda$ and $\Gamma$. Here we considered $m=0.5$ and $\mu=0.1$. The
intersection with the plane
$F=1$ occurs for $\lambda$ below the
critical value $\lambda_{c}^{(1)}=1.048$. We might emphasize here that we took $\alpha=1$, but the qualitative behaviour of the gluon propagator does not depend on $\alpha$.}
\end{figure}

In this part, we will delve into a more intricate numerical exploration of the theory's regimes and pinpoint the critical temperatures pertinent to these phase transitions. 
Our approach involves a numerical scrutiny of the gap equation (\ref{gap equation RGZ}). In essence, we will illustrate the dependence of the left-hand side of eq. (\ref{gap equation RGZ}) on the parameters $\lambda$, which embodies the temperature dependence, and $\Gamma$, which embodies the Gribov parameter dependence, as demonstrated in Figure [\ref{fig:1}], where
\begin{eqnarray}
F(\lambda,\Gamma,m,\mu)=\frac{3Ng^{2}\lambda}{16\pi^{3}}\int_{0}^{1}{dR}\Bigg(R^{2}S(R,\lambda,\Gamma,m,\mu)\Bigg),\nonumber\\
\end{eqnarray}
with the coupling constant $g$ being given in the hard thermal regime ($T>>1$) by
\begin{eqnarray}
g^{2}(\lambda)=\frac{8\pi^{2}}{11\ln{\bigg(\frac{2\pi T}{\Lambda_{QCD}}}\bigg)}=\frac{8\pi^{2}}{11\ln{(\alpha\lambda)}},\label{coupling constant1}
\end{eqnarray}
and 
\begin{eqnarray}
\alpha\equiv \frac{\Lambda}{\Lambda_{QCD}}.
\end{eqnarray}
In this illustration, it is evident that the existence of the gap equation's solution, denoted by the plane $F(\lambda,\Gamma,m,\mu)=1$, depends ultimately on the temperature $T$.

\begin{figure}[h!]
\centering
\includegraphics[width=0.5\textwidth]{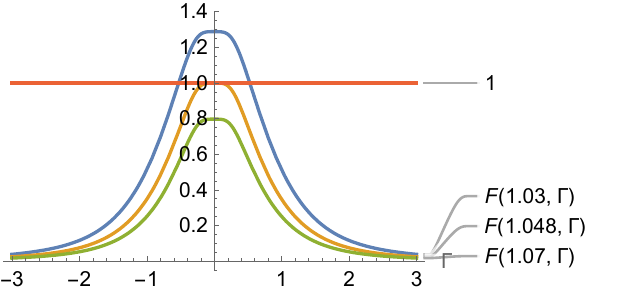}
\caption{\label{fig:2}Plot of the surface $F(\lambda,\Gamma)$ when $\lambda=1.03$, $\lambda=1.048$, $\lambda=1.07$. Here we considered $m=0.5$ and $\mu=0.1$. The red line is where plane $F=1$ intercept the blue surface of figure [\ref{fig:1}].}
\end{figure}

\begin{figure}[h!]
\centering
\includegraphics[width=0.52
\textwidth]{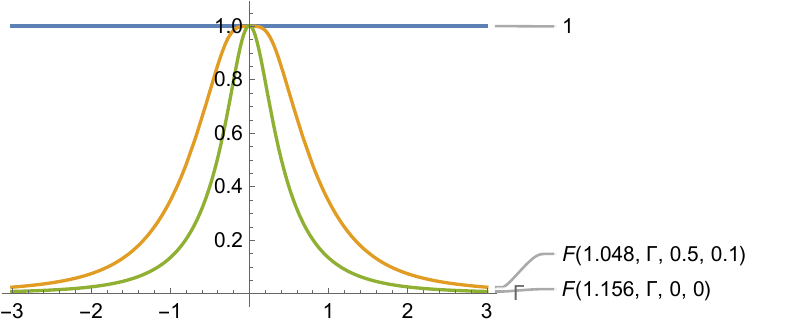}
\caption{\label{fig:2.1}Comparison of the plot of the surface $F(\lambda,\Gamma,m,\mu)$ for the case when the masses of the condensates are $m=0.5$ and $\mu=0.1$ with the case in which they are null.}
\end{figure}

\begin{figure}[h!]
\centering
\includegraphics[width=0.4\textwidth]{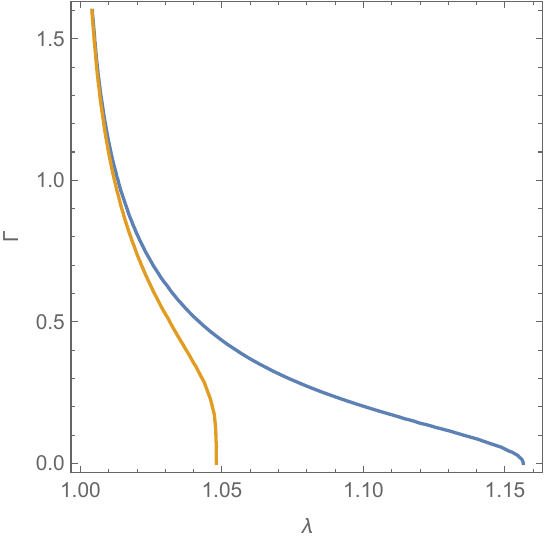}
\caption{\label{fig:3}Plot of $\Gamma(\lambda)$. Here we considered $m=0.5$ and $\mu=0.1$ for the orange curve. It can be seen that $\Gamma$ reaches zero when $\lambda$ reaches the critical value $\lambda_{c}^{(1)}=1.048$. Yet the blue curve represents the case in which the masses of the condensates are null. For this case $\Gamma$ reaches zero when $\lambda$ reaches the critical value $\lambda_{c}^{(1)}=1.156$. }
\end{figure}

\begin{figure}[h!]
\centering
\includegraphics[width=0.5\textwidth]{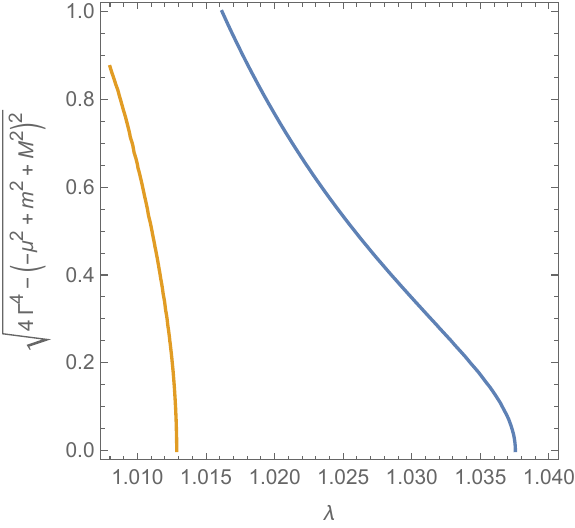}
\caption{\label{fig:4}Plot of $\sqrt{4\Gamma^{4}-\left(-\mu ^2+m^2+M^2\right)^2}$ in terms of $\lambda$. We can see here that when $m=0.5$ and $\mu=0.1$ there is a phase transition at $\lambda_{c}^{(2)}=1.0128$ from a semi-confined to a confined phase. In other words, we can say that the critical temperature of such a phase transition is $\frac{T_{c}^{(2)}}{\Lambda}=0.1612$. However, when the masses of the condensates are turned off, the phase transition occurs at $\lambda_{c}^{(2)}=1.0375$ and the critical temperature becomes $\frac{T_{c}^{(2)}}{\Lambda}=0.1651$}
\end{figure}

From figure [\ref{fig:2}], it is evident that the critical temperature is reached when $\lambda_{c}^{(1)}=1.048$. This corresponds to
\begin{eqnarray}
\frac{T_{c}^{(1)}}{\Lambda_{QCD}}=0.1668,
\end{eqnarray}
assuming the condensate masses are held constant at $m=0.5$ and $\mu=0.1$. When these massive terms are disregarded, as shown in figure [\ref{fig:2.1}], the critical temperature shifts to
\begin{eqnarray}
\frac{T_{c}^{(1)}}{\Lambda_{QCD}}=0.1840,
\end{eqnarray}
This indicates that the presence of condensate masses contributes to a decrease in the phase transition temperature.


Then the three regimes can be interpreted as follows:
\begin{itemize}
    \item For $T>T_{c}^{(1)}$ (or $\lambda>\lambda_{c}^{(1)}$), the gap equation does not have a solution, implying that the massive Gribov parameter is zero and all gluonic degrees of freedom are asymptotically free in this regime.
    \item For $T<T_{c}^{(1)}$, solutions to the gap equation emerge, thereby defining the Gribov parameter $\gamma$. In this regime, as depicted in figure [\ref{fig:3}], the $\Gamma$ parameter diminishes to zero as the temperature rises. This occurs when $\lambda_{c}^{(1)}=1.048$ in the RGZ case and when $\lambda_{c}^{(1)}=1.1840$ in the GZ case. Despite the existence of a solution for the gap equation, total confinement of the propagator only occurs when the discriminant $4\Gamma^{4}-\left(-\mu ^2+m^2+M^2\right)^2$ of eq. (\ref{polos}) alters its sign. As illustrated in figure [\ref{fig:4}] and eq. (\ref{polos}), for $\lambda_{c}^{1}>\lambda>\lambda_{c}^{2}$, with $\lambda_{c}^{2}=1.0128$ for the RGZ case and $\lambda_{c}^{(2)}=1.0375$ for the GZ case, a partially confined phase exists, with one confined and one deconfined (physical) degrees of freedom.
   
    \item For $T<T_{c}^{2}$, with $\frac{T_{c}^{(2)}}{\Lambda_{QCD}}=0.1612$, a completely confined phase is observed in the RGZ case, while the same occurs at $\frac{T_{c}^{(2)}}{\Lambda_{QCD}}=0.1651$ in the GZ case.
\end{itemize}

It is crucial to note that qualitatively, our findings align perfectly with those presented in \cite{canfora2014gribov}, where only the Gribov copies in the Landau gauge were considered. The sole distinction in our results is the lower critical temperatures, attributable to our consideration of the masses $m$ and $\mu$.

\section{The infrared regime}\label{section11}
After having examined the hard thermal regime, it is now necessary to make a similar analysis for the infrared regime, characterized by $\lambda<1$. To do so, it is necessary to extend of the coupling constant $g$ to the infrared regime, which is accomplished consistently by
\begin{eqnarray}
g^{2}(g_{0},\lambda)=\frac{g_{0}^{2}}{1+\frac{11}{16\pi^{2}}g_{0}^{2}\ln{(1+\alpha^{2}\lambda^{2})}},\label{g0}
\end{eqnarray}
as the mass $M^{2}$ approaches zero and the coupling constant $g$ goes to $g_{0}$ when the temperature $T$ also reduces to zero, allowing us to reobtain the zero-temperature results.

%

\begin{figure}[h!]
\centering
\includegraphics[width=0.5\textwidth]{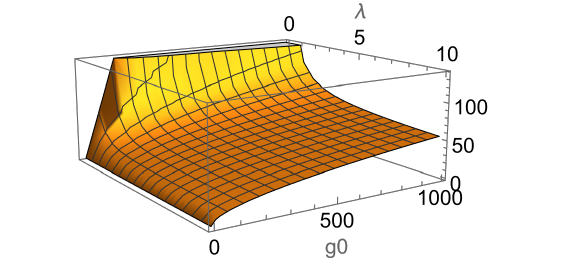}
\caption{\label{fig:5}Plot of $g(g_{0},\lambda)$.}
\end{figure}

It should be noted that for large values of $g_{0}$, the behaviour of $g(g_{0}, \lambda)$ becomes insensitive to minor variations in $g_{0}$ itself, as illustrated in figure [\ref{fig:5}]. This aligns with the principle in quantum field theory that bare quantities are infinite yet unobservable, necessitating their renormalization.

\begin{figure}[h!]
\centering
\includegraphics[width=0.5\textwidth]{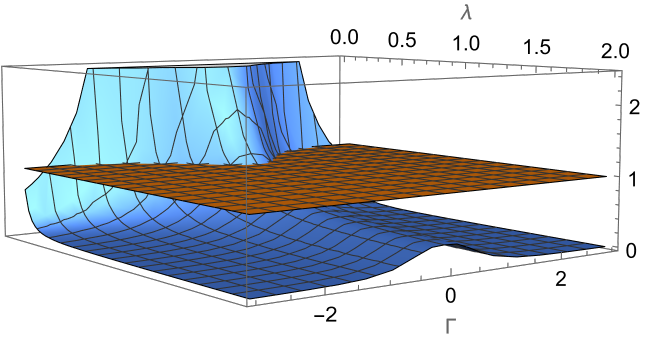}
\caption{\label{fig:6}Plot of the surface $G$ for
different values of $\lambda$ and $\Gamma$. The
intersection with the plane
$G=1$ occurs for $\lambda$ below the
critical value $\lambda_{c}^{(1)}=0.628$.}
\end{figure}

Similar to the analysis in section (\ref{section9}), we will numerically examine the gap equation, which in this context is expressed as
\begin{eqnarray}
G(g_{0},\lambda,\Gamma)=\frac{3g^{2}N\lambda}{16\pi^{3}}\int_{0}^{1}dR(R^{2}S(R,g_{0},\lambda,\Gamma)).
\end{eqnarray}
Upon selecting $g_{0}=1000$, we observe a qualitative behaviour identical to that found in section \ref{section9}, with the same phase transitions, as depicted in figure [\ref{fig:6}].
%
%
%
%
%
%
%
%
%
%
%
%
%
%
%
%

\begin{figure}[h!]
\centering
\includegraphics[width=0.5\textwidth]{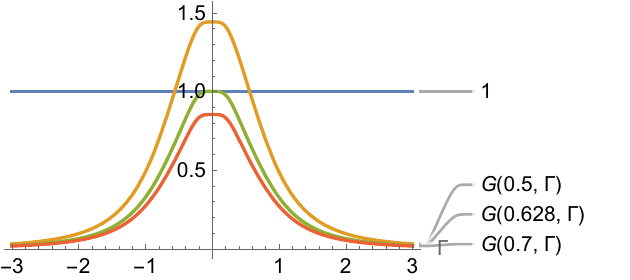}
\caption{\label{fig:7}Plot of the surface $G(\lambda,\Gamma,m,\mu)$ when $\lambda=0.5$, $\lambda=0.628$, $\lambda=0.7$ (for $m=0.5$ and $\mu=0.1$). Here the blue line is where plane $G=1$ intercept the surface of figure [\ref{fig:6}].}
\end{figure}

\begin{figure}[h!]
\centering
\includegraphics[width=0.54\textwidth]{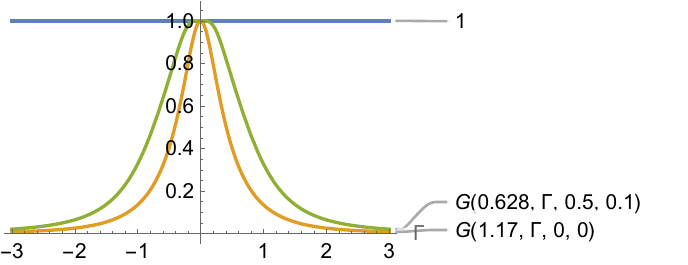}
\caption{\label{fig:7.1}Plot of the surface $G(\lambda,\Gamma,m,\mu)$ when $\lambda=0.628$ (for $m=0.5$ and $\mu=0.1$), and $\lambda=1.17$ (for $m=0$ and $\mu=0$).}
\end{figure}

\begin{figure}[h!]
\centering
\includegraphics[width=0.4\textwidth]{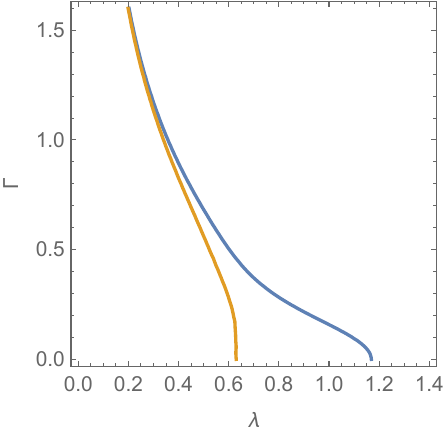}
\caption{\label{fig:8}Plot of $\Gamma(\lambda)$. Here we can see that $\Gamma$ reaches zero when $\lambda$ reaches the critical value $\lambda_{c}^{(1)}=0.628$ for $m=0.5$ and $\mu=0.1$. Yet for $m=\mu=0$ it occurs at $\lambda_{c}^{(1)}=1.17$}
\end{figure}

\begin{figure}[h!]
\centering
\includegraphics[width=0.4\textwidth]{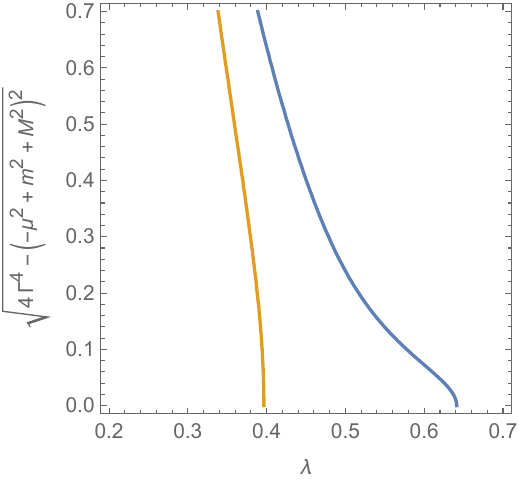}
\caption{\label{fig:9}Plot of $\sqrt{4\Gamma^{4}-(-\mu^{2}+m^{2}+M^{2})^{2};}$ in terms of $\lambda$. We can see here that for $m=0.5$ and $\mu=0.1$ there is a phase transition at $\lambda_{c}^{(2)}=0.396$ from a semi-confined to a confined phase. In other words, we can say that the critical temperature of such a phase transition is $\frac{T_{c}^{(2)}}{\Lambda}=0.063$. However, for $m=\mu=0$ the phase transition happens at $\lambda_{c}^{(2)}=0.64$, which means at the critical temperature $\frac{T_{c}^{(2)}}{\Lambda}=0.102$.}
\end{figure}

The critical temperatures for this instance are determined as follows:
\begin{itemize}
    \item For the phase transition from the deconfined to the semi-confined regime, we have
    \begin{eqnarray}
    \lambda_{c}^{(1)}=0.628\rightarrow \frac{T_{c}^{(1)}}{\Lambda_{QCD}}=0.0999\approx 0.1,
    \end{eqnarray}
    given $m=0.5$ and $\mu=0.1$. Conversely, when $m=\mu=0$, we find
    \begin{eqnarray}
    \lambda_{c}^{(1)}=1.17\rightarrow \frac{T_{c}^{(1)}}{\Lambda_{QCD}}=0.1862\approx 0.1,
    \end{eqnarray}
    as depicted in figures [\ref{fig:7}], [\ref{fig:7.1}], and [\ref{fig:8}].
    \item For the transition from the semi-confined to the confined regime, we have
    \begin{eqnarray}
    \lambda_{c}^{(2)}=0.396\rightarrow \frac{T_{c}^{(2)}}{\Lambda_{QCD}}=0.063,
    \end{eqnarray}
    when $m=0.5$ and $\mu=0.1$. However, when $m=\mu=0$, it becomes
     \begin{eqnarray}
    \lambda_{c}^{(2)}=0.64\rightarrow \frac{T_{c}^{(2)}}{\Lambda_{QCD}}=0.102,
    \end{eqnarray}
    as shown in figure [\ref{fig:9}].
\end{itemize}
%
%
%
%
%
%
Before concluding, we must emphasize that the qualitative behaviour of our plots do not depend on $\alpha$. The only thing it would change is that increasing/decreasing the $\alpha$ value, it would increase/decrease the critical tempertures in a completely analogous way to the one showed on \cite{canfora2014gribov}.

\section{Calculating the thermal mass in terms of the RGZ masses}\label{section12}
It is crucial to note that our approach thus far has involved the introduction of a term of the form $\frac{M^{2}}{2}A_{\mu}^{a}A_{\mu}^{a}$ into the theory's Lagrangian. We then computed its impact on the theory's physical spectrum, specifically in relation to inducing a phase transition within the theory.

However, it should be observed that the thermal mass $M^{2}$ we derived was given in terms of the components $\Pi_{44}^{(2)}$ and $\Pi_{\mu\mu}^{(2)}$ of the gluon self-energy. These were computed without taking into account the presence of Gribov copies and the masses of the condensates.

Our objective now is to perform an analogous calculation, but this time incorporating diagrams that represent the $2$-point functions of Yang-Mills theory within the framework of Refined Gribov-Zwanziger at $1$-loop.




\subsection{Diagrams}
The Refined Gribov-Zwanziger action at $1$-loop (\ref{S RGZ}) generates $13$ diagrams. However, under the High Thermal Limit (HTL), only $7$ of these diagrams contribute to the gluon self-energy. The remaining $6$ diagrams become negligible when the internal momentum $k$ approaches zero.



These $13$ diagrams are,\\
	\begin{tikzpicture}
		\begin{feynman}
			\vertex (a);
			\vertex [right=0.5cm of a] (b);
			\vertex [right=0.5cm of b] (c);
			\vertex [right=0.5cm of c] (d);
			\vertex [right=0.15cm of d] (e) {$=$};
			\vertex [right=0.25cm of e] (f);
			\vertex [right=0.5cm of f] (g);
			\vertex [right=0.5cm of g] (h);
			\vertex [right=0.5cm of h] (i);
			\vertex [right=0.15cm of i] (j) {$+$};
			\vertex [right=0.25cm of j] (k);
			\vertex [right=0.75cm of k] (l);
                \vertex [below=0.15cm of l] (l1) {$(b)$};
			\vertex [above=0.5cm of l] (m);
			\vertex [right=0.75cm of l] (n);
			\vertex [right=0.15cm of n] (o) {$+$};
			\vertex [right=0.25cm of o] (p);
			\vertex [right=0.5cm of p] (q);
			\vertex [right=0.5cm of q] (r);
			\vertex [right=0.5cm of r] (s);
			\vertex [below=of e] (t) {$+$};
			\vertex [right=0.25cm of t] (u);
			\vertex [right=0.5cm of u] (v);
			\vertex [right=0.5cm of v] (x);
			\vertex [right=0.5cm of x] (y);
			\vertex [right=0.15cm of y] (w) {$+$};
			\vertex [right=0.25cm of w] (a1);
			\vertex [right=0.5cm of a1] (b1);
			\vertex [right=0.5cm of b1] (c1);
			\vertex [right=0.5cm of c1] (d1);
			\vertex [right=0.15cm of d1] (e1) {$+$};
			\vertex [right=0.25cm of e1] (f1);
			\vertex [right=0.5cm of f1] (g1);
			\vertex [right=0.5cm of g1] (h1);
			\vertex [right=0.5cm of h1] (i1);
			\vertex [below=of t] (j1) {$+$};
			\vertex [right=0.25cm of j1] (k1);
			\vertex [right=0.25cm of k1] (l1);
			\vertex [right=0.25cm of l1] (m1);
			\vertex [right=0.5cm of m1] (n1);
			\vertex [right=0.25cm of n1] (o1);
			\vertex [right=0.25cm of o1] (p1);
			\vertex [right=0.15cm of p1] (q1) {$+$};
			\vertex [right=0.15cm of q1] (r1);
			\vertex [right=0.25cm of r1] (s1);
			\vertex [right=0.25cm of s1] (t1);
			\vertex [right=0.5cm of t1] (u1);
			\vertex [right=0.25cm of u1] (v1);
			\vertex [right=0.25cm of v1] (x1);
			\vertex [right=0.15cm of x1] (y1) {$+$};
			\vertex [right=0.25cm of y1] (a2);
			\vertex [right=0.25cm of a2] (b2);
			\vertex [right=0.25cm of b2] (c2);
			\vertex [above right=0.35cm of c2] (d2);
			\vertex [below right=0.35cm of d2] (e2);
			\vertex [right=0.5cm of e2] (f2);
			\vertex [below=of j1] (g2) {$+$};
			\vertex [right=0.25cm of g2] (h2);
			\vertex [right=0.5cm of h2] (i2);
			\vertex [above right=0.35cm of i2] (j2);
			\vertex [below right=0.35cm of j2] (k2);
			\vertex [below right=0.35cm of i2] (l2);
                \vertex [below=0.001cm of l2] (ll) {$(j)$}; 
			\vertex [right=0.5cm of k2] (m2);
			\vertex [right=0.15cm of m2] (n2) {$+$};
			\vertex [right=0.25cm of n2] (o2);
			\vertex [right=0.25cm of o2] (p2);
			\vertex [right=0.25cm of p2] (q2);
			\vertex [above right=0.35cm of q2] (r2);
			\vertex [below right=0.35cm of r2] (s2);
			\vertex [below right=0.35cm of q2] (t2);
                \vertex [below=0.001cm of t2] (tt) {$(k)$}; 
			\vertex [right=0.25cm of s2] (u2);
			\vertex [right=0.25cm of u2] (v2);
			\vertex [right=0.15cm of v2] (x2) {$+$};
			\vertex [right=0.25cm of x2] (a3);
			\vertex [right=0.25cm of a3] (b3);
			\vertex [right=0.25cm of b3] (c3);
			\vertex [below right=0.35cm of c3] (d3);
			\vertex [above right=0.35cm of d3] (e3);
                \vertex [below=0.001cm of d3] (dd) {$(l)$}; 
			\vertex [right=0.5cm of e3] (f3);
			\vertex [below=of g2] (g3) {$+$};
			\vertex [right=0.25cm of g3] (h3);
			\vertex [right=0.25cm of h3] (i3);
			\vertex [right=0.25cm of i3] (j3);
			\vertex [below right=0.34cm of j3] (k3);
                \vertex [below=0.001cm of k3] (kk) {$(m)$}; 
			\vertex [above right=0.34cm of k3] (l3);
			\vertex [right=0.5cm of l3] (m3);

			
			\diagram{
				(a) -- [boson] (b);
				(b) -- [ half left, fill=gray] (c);
				(c) -- [ half left, fill=gray] (b);
				(c) -- [boson] (d);
				
				(f) -- [boson] (g);
				(g) -- [ghost, half left] (h);
				(h) -- [ghost, half left, edge label=$(a)$] (g);
				(h) -- [boson] (i);
				
				(k) -- [boson] (l);
				(l) -- [boson, half left] (m);
				(m) -- [boson, half left] (l);
				(l) -- [boson] (n);
				
				(p) -- [boson] (q);
				(q) -- [boson, half left] (r);
				(r) -- [boson, half left, edge label=$(c)$] (q);
				(r) -- [boson] (s);
				
				(u) -- [boson] (v);
				(v) -- [plain, half left] (x);
				(x) -- [plain, half left, edge label=$(d)$] (v);
				(x) -- [boson] (y);
				
				(a1) -- [boson] (b1);
				(b1) -- [double, half left] (c1);
				(c1) -- [double, half left, edge label=$(e)$] (b1);
				(c1) -- [boson] (d1);
				
				(f1) -- [boson] (g1);
				(g1) --[scalar, half left] (h1);
				(h1) --[scalar, half left, edge label=$(f)$] (g1);
				(h1) -- [boson] (i1);
				
				(k1) -- [boson] (l1);
				(l1) -- [double] (m1);
				(m1) -- [plain, half left] (n1);
				(n1) -- [boson, half left, edge label=$(g)$] (m1);
				(n1) -- [double] (o1);
				(o1) -- [boson] (p1);
				
				(r1) -- [boson] (s1);
				(s1) -- [double] (t1);
				(t1) -- [double, half left] (u1);
				(u1) -- [boson, half left, edge label=$(h)$] (t1);
				(u1) -- [double] (v1);
				(v1) -- [boson] (x1);
				
				(a2) -- [boson] (b2);
				(b2) -- [double] (c2);
				(c2) -- [double, quarter left] (d2);
				(d2) -- [boson, quarter left] (e2);
				(e2) -- [boson, half left, edge label=$(i)$] (c2);
				(e2) -- [boson] (f2);
				
				(h2) -- [boson] (i2);
				(i2) -- [double, quarter left] (j2);
				(j2) -- [boson, quarter left] (k2);
				(i2) -- [double, quarter right] (l2);
				(l2) -- [boson, quarter right] (k2);
				(k2) -- [boson] (m2);
				
				(o2) -- [boson] (p2);
				(p2) -- [double] (q2);
				(q2) -- [double, quarter left] (r2);
				(r2) -- [boson, quarter left] (s2);
				(q2) -- [boson, quarter right] (t2);
				(t2) -- [double, quarter right] (s2);
				(s2) -- [double] (u2);
				(u2) -- [boson] (v2);
				
				(a3) -- [boson] (b3);
				(b3) -- [double] (c3);
				(c3) -- [plain, half left] (e3);
				(c3) -- [boson, quarter right] (d3);
				(d3) -- [double, quarter right] (e3);
				(e3) -- [boson] (f3);
				
				(h3) -- [boson] (i3);
				(i3) -- [double] (j3);
				(j3) -- [boson, quarter right] (k3);
				(k3) -- [double, quarter right] (l3);
				(j3) -- [double, half left] (l3);
				(l3) -- [boson] (m3);
			};

		\end{feynman}

	\end{tikzpicture}
where,\\\\
       %
       %
       %
    %
%
 %
%
%
%
\begin{tikzpicture}
    \begin{feynman}
        \vertex (a);
        \vertex [right=of a] (b) {$(i)$};
        \vertex [right=1cm of b] (c);
        \vertex [right=of c] (d) {$(ii)$};
        \vertex [right=1cm of d] (e);
        \vertex [right=of e] (f) {$(iii)$};
        \vertex [below=1cm of a] (o);
        \vertex [right=of o] (p) {$(iv)$};
        \vertex [right=1cm of p] (g);
        \vertex [right=of g] (h) {$(v)$};
        \vertex [right=1cm of h] (i);
        \vertex [right=0.75cm of i] (j);
        \vertex [right=0.75cm of j] (k) {$(vi)$};
        \vertex [below=1cm of i] (l);
        \vertex [right=0.75cm of l] (m);

        \diagram{
        (a) -- [boson] (b);
        (c) -- [ghost] (d);
        (e) -- [double] (f);
        (o) -- [plain] (p);
        (g) -- [scalar] (h);
        (i) -- [double] (j) -- [boson] (k);

        };
    \end{feynman}
\end{tikzpicture}\\

are respectively the gluon propagator $(i)$,
\begin{eqnarray}
\langle A_{\mu}^{a}(k)A_{\nu}^{b}(-k)\rangle&=&\Bigg[\frac{k^{2}+\mu^{2}}{((k^{2}+\mu^{2})(k^{2}+m^{2})+2g^{2}N\gamma^{4})}\times\nonumber\\
&\times&\Bigg(\delta_{\mu\nu}-\frac{k_{\mu}k_{\nu}}{k^{2}}\Bigg)+\alpha\frac{k_{\mu}k_{\nu}}{k^{4}}\Bigg]\delta^{ab},\nonumber\label{prop gluon lin cov rgz}
\end{eqnarray}
the ghost propagator $(ii)$,
\begin{eqnarray}
\langle \bar{c}^{a}(k)c^{b}(-k)\rangle=-\frac{\delta^{ab}}{k^{2}}.
\end{eqnarray}
the bosonic auxiliary field propagator $(iii)$,
\begin{eqnarray}
\langle \bar{\phi}_{\mu}^{ab}(k)\phi_{\nu}^{cd}(-k)\rangle&=&\frac{-g^{2}\gamma^{4}f^{abr}f^{cdr}}{((k^{2}+\mu^{2})(k^{2}+m^{2})+2g^{2}N\gamma^{4})}\times\nonumber\\
&\times&\frac{1}{(k^{2}+\mu^{2})}\Bigg(\delta_{\mu\nu}-\frac{k_{\mu}k_{\nu}}{k^{2}}\Bigg)+\frac{\delta^{ac}\delta^{bd}\delta_{\mu\nu}}{k^{2}+\mu^{2}},\nonumber\\
\end{eqnarray}
the bosonic auxiliary field propagator $(iv)$,
\begin{eqnarray}
\langle \phi_{\mu}^{ab}(k)\phi_{\nu}^{cd}(-k)\rangle&=&\frac{-g^{2}\gamma^{4}f^{abr}f^{cdr}}{((k^{2}+\mu^{2})(k^{2}+m^{2})+2g^{2}N\gamma^{4})}\times\nonumber\\
&\times&\frac{1}{(k^{2}+\mu^{2})}\Bigg(\delta_{\mu\nu}-\frac{k_{\mu}k_{\nu}}{k^{2}}\Bigg),\nonumber\\
\langle \phi_{\mu}^{ab}(k)\phi_{\nu}^{cd}(-k)\rangle&=&\langle \bar{\phi}_{\mu}^{ab}(k)\bar{\phi}_{\nu}^{cd}(-k)\rangle,
\end{eqnarray}
the fermionic auxiliary field propagator $(v)$,
\begin{eqnarray}
\langle \bar{\omega}_{\mu}^{ab}(k)\omega_{\nu}^{cd}(-k)\rangle=-\frac{\delta^{ac}\delta^{bd}\delta_{\mu\nu}}{k^{2}+\mu^{2}}
\end{eqnarray}
and the mixed propagator $(vi)$,
\begin{eqnarray}
\langle \phi_{\mu}^{ab}(k)A_{\nu}^{c}(-k)\rangle&=&-\frac{ig\gamma^{2}f^{abc}}{((k^{2}+\mu^{2})(k^{2}+m^{2})+2g^{2}N\gamma^{4})}\times\nonumber\\
&\times&\Bigg(\delta_{\mu\nu}-\frac{k_{\mu}k_{\nu}}{k^{2}}\Bigg),\nonumber\\
\langle \phi_{\mu}^{ab}(k)A_{\nu}^{c}(-k)\rangle&=&\langle \bar{\phi}_{\mu}^{ab}(k)A_{\nu}^{c}(-k)\rangle.
\end{eqnarray}

Concerning the vertices of the theory, we have that besides (\ref{vert123}), we will also have the following ones,
\begin{eqnarray}
    \left[V_{A\bar{\phi}\phi}(k,p,q)\right]_{\mu\nu\rho}^{abcde}&=&-(2\pi)^{4}igf^{abd}\delta^{ce}\delta_{\mu\nu}p_{\rho},\\
    \left[V_{A\bar{\omega}\omega}(k,p,q)\right]_{\mu\nu\rho}^{abcde}&=&-(2\pi)^{4}igf^{abd}\delta^{ce}\delta_{\mu\nu}p_{\rho},
\end{eqnarray}

Finally, we can build the diagrams that will survive the HTL limit $k\rightarrow0$.
The first one is the diagram $(a)$,
\begin{eqnarray}
i\Pi_{\mu\nu}^{(2)ab}(k)&=&\int{\frac{d^{d}p}{(2\pi)^{d}}}\Bigg(\frac{g^2 N p^{\mu } p^{\nu } \delta^{ab}}{p^2 \left(k-p\right)^{2}}\Bigg).\label{1}
\end{eqnarray}
The second one is the diagram $(b)$ is,
\begin{eqnarray}
i\Pi_{\mu\nu}^{(2)ab}(k)&=&\int{\frac{d^{d}p}{(2\pi)^{d}}}\Bigg(-\frac{2 g^2 N \left(m^2-p^2\right) \delta ^{ab}}{p^2}\nonumber\\
   &\times&\frac{\left(p^2 (\alpha +\mathit{d}-2) \delta^{\mu \nu }-(\alpha -1) p^{\mu } p^{\nu }\right)}{ \left(2 \gamma ^4 g^2 N+\mu ^2 m^2-p^2 \left(\mu ^2+m^2\right)+\left(p^2\right)^2\right)}\Bigg).\label{2}\nonumber\\
\end{eqnarray}
The third diagram $(c)$ is,
\begin{eqnarray}
i\Pi_{\mu\nu}^{(2)ab}(k)&=&\int{\frac{d^{d}p}{(2\pi)^{d}}}\Bigg(-\frac{2 g^2 N \left(m^2-p^2\right)^2 \delta ^{ab} }{\left(p^2-x_{1}^{2}\right) \left(p^2-x_{2}^{2}\right) }\nonumber\\
   &\times&\frac{\left((-\alpha +2 \mathit{d}-2) p^{\mu } p^{\nu }+\alpha  p^2 g^{\mu \nu
   }\right)}{\left((k-p)^{2}-x_{1}^{2}\right) \left((k-p)^{2}-x_{2}^{2}\right)} \Bigg).\label{3}
\end{eqnarray}
The fourth diagram $(d)$ is,
\begin{eqnarray}
i\Pi_{\mu\nu}^{(2)ab}(k)&=&\int{\frac{d^{d}p}{(2\pi)^{d}}}\Bigg(-\frac{\gamma ^8 (\mathit{d}-1) g^6 N^3 p^{\mu } p^{\nu } \delta ^{ab}}{4 \left(p^2-m^2\right) \left(x_{1}^{2}-p^2\right)
   \left(p^2-x_{2}^{2}\right)}\nonumber\\
   &\times&\frac{1}{ \left((k-p)^{2}-m^2\right) \left(-(k-p)^{2}+x_{1}^{2}\right)}\nonumber\\
   &\times&\frac{1}{\left((k-p)^{2}-x_{2}^{2}\right)}\Bigg).\label{4}
\end{eqnarray}
The fifth diagram $(e)$ is,
\begin{eqnarray}
i\Pi_{\mu\nu}^{(2)ab}(k)&=&\int{\frac{d^{d}p}{(2\pi)^{d}}}\Bigg\{
g^2 N^2 p^{\mu } p^{\nu } \delta ^{ab} \Bigg[8 p^2 \left(-p^2+x_{1}^{2}+x_{2}^{2}\right)\nonumber\\
&\times&\Bigg(\mathit{d} p^2 \frac{(N^{2}-1)}{2N}
   \left(p^2-x_{1}^{2}-x_{2}^{2}\right)+\nonumber\\
   &+&2 \mathit{d} x_{1}^{2} x_{2}^{2} \frac{(N^{2}-1)}{2N}-\left(\gamma ^4 (\mathit{d}-1) g^2\right)\Bigg)+\nonumber\\
   &-&8\mathit{d} x_{1}^{4} x_{2}^{4} \frac{(N^{2}-1)}{2N}-\left(\gamma ^8 (\mathit{d}-1) g^4 N\right)+\nonumber\\
   &+&8 \gamma ^4 (\mathit{d}-1) g^2 x_{1}^{2}
   x_{2}^{2}\Bigg]\nonumber\\
   &\times&\frac{1}{4 \left(m^2-p^2\right) \left(p^2-x_{1}^{2}\right) \left(p^2-x_{2}^{2}\right) \left((k-p)^{2}-x_{1}^{2}\right)}\nonumber\\
   &\times&\frac{1}{\left(-(k-p)^{2}+m^2\right)
    \left((k-p)^{2}-x_{2}^{2}\right)}\Bigg\}.\label{7}\nonumber\\
\end{eqnarray}
The sixth diagram $(f)$ is,
\begin{eqnarray}
i\Pi_{\mu\nu}^{(2)ab}(k)&=&\int\frac{d^{d}p}{(2\pi)^{d}}\Bigg(\frac{\mathit{d} g^2 N \left(N^2-1\right) p^{\mu } p^{\nu } \delta ^{ab}}{\left(m^2-p^2\right) \left(-(k-p)^{2}+m^2\right)}\Bigg).\label{5}\nonumber\\
\end{eqnarray}
The seventh diagram mixed $(j)$ is,
\begin{eqnarray}
i\Pi_{\mu\nu}^{(2)ab}(k)&=&\int{\frac{d^{d}p}{(2\pi)^{d}}}\Bigg(\frac{\gamma ^4 (\mathit{d}-1) g^4 N^2 p^{\mu } p^{\nu } \delta ^{ab}}{\left(x_{1}^{2}-p^2\right) \left(p^2-x_{2}^{2}\right)}\nonumber\\
&\times&\frac{1}{ \left(-(k-p)^{2}+x_{1}^{2}\right) \left((k-p)^{2}-x_{2}^{2}\right)}\Bigg),\label{6}
\end{eqnarray}
where $x_{1}^{2}$ and $x_{2}^{2}$ are given by,
\begin{eqnarray}
x_{1}^{2}&=&\frac{1}{2} \left(\sqrt{-8 \gamma ^4 g^2 N+m^4-2 m^2 M^2+M^4}+m^2+M^2\right),\nonumber\\
x_{2}^{2}&=&\frac{1}{2} \left(-\sqrt{-8 \gamma ^4 g^2 N+m^4-2 m^2 M^2+M^4}+m^2+M^2\right).\nonumber\\
\end{eqnarray}

Summing the diagrams \ref{1}, \ref{2}, \ref{3}, \ref{4}, \ref{7}, \ref{5} and \ref{6} we obtain the gluon self-energy $i\Pi_{\mu\nu}^{(2)ab}$. By considering the components $i\Pi_{44}^{(2)}$ and $i\Pi_{\mu\mu}^{(2)}$, we derive both the longitudinal and transverse thermal masses, denoted as $\Pi_{L}$ and $\Pi_{T}$ respectively, utilizing eqs. (\ref{regras1}) and (\ref{regras2}). 
Then taking into account the $0$-temperature propagator (\ref{prop RGZ lin}), it can be rewritten as follows,
\begin{eqnarray}
 \langle A_{\mu}^{a}(k)A_{\nu}^{b}(-k) \rangle=\Bigg[\frac{1}{(k^{2}+\bar{m}^{2})}\Bigg(\delta_{\mu\nu}-\frac{k_{\mu}k_{\nu}}{k^{2}}\Bigg)+\frac{\alpha k_{\mu}k_{\nu}}{k^{4}}\Bigg]\delta^{ab},\nonumber\\
\end{eqnarray}
where,
\begin{eqnarray}
\bar{m}^{2}=m^{2}+\frac{2Ng_{0}^{2}\gamma^{4}}{k^{2}+\mu^{2}},
\end{eqnarray}
This implies that upon performing the analytic continuation $i\omega\rightarrow\omega+i\delta$, with the Minkowski four-momentum, $K_{\mu}=(\omega,\vec{k})$, the propagator of the theory at finite temperatures becomes,
\begin{eqnarray}
D_{\mu\nu}(K)&=&-\frac{(P_{T})_{\mu\nu}}{K^{2}-\bar{m}^{2}-\Pi_{T}}-\frac{(P_{L})_{\mu\nu}}{K^{2}-\bar{m}^{2}-\Pi_{L}}-\alpha\frac{K_{\mu}K_{\nu}}{K^{4}},\nonumber\\
\bar{m}^{2}&=&m^{2}-\frac{2Ng_{0}^{2}\gamma^{4}}{K^{2}-\mu^2}, \label{prop finite t RGZ2}
\end{eqnarray}
or, in other words,
\begin{eqnarray}
D_{\mu\nu}(K)&=& \frac{\left(-K^2+\mu^2 \right)(P_T)_{ \mu\nu }} {(-K^2+m^2+\Pi_T)(-K^2+\mu^2)+ 2Ng_0^2 \gamma^4 } + \nonumber\\
&+& \frac{\left(-K^2+\mu^2 \right)(P_L)_{ \mu\nu }} {(-K^2+m^2+\Pi_L)(-K^2+\mu^2)+ 2Ng_0^2 \gamma^4 } + \nonumber\\
&-& \alpha \frac { K_\mu K_\nu } { K^4 }, \label { prop final }
\end{eqnarray}
where $\Pi_L$ and $\Pi_T$ are intricate functions of $m$, $\mu$, and $\gamma$, even after all divergent terms have been removed.


Therefore, a significant finding that can be seen now is that the masses $\Pi_{L}$ and $\Pi_{T}$ do not coincide in the limit as $k\rightarrow0$, since they are given by,
\begin{widetext}
\begin{eqnarray}
x_{1}^{2}&=&\frac{1}{2}\left(\mu^{2}+m^{2}+\Pi_{L}-\sqrt{-8Ng^{2}\gamma^{4}+\mu^{4}-2\mu^{2}m^{2}+m^{4}-2\mu^{2}\Pi_{L}+2m^{2}\Pi_{L}+\Pi_{L}^{2}}\right),\nonumber\\
x_{2}^{2}&=&\frac{1}{2}\left(\mu^{2}+m^{2}+\Pi_{L}+\sqrt{-8Ng^{2}\gamma^{4}+\mu^{4}-2\mu^{2}m^{2}+m^{4}-2\mu^{2}\Pi_{L}+2m^{2}\Pi_{L}+\Pi_{L}^{2}}\right),
\end{eqnarray}
\end{widetext}
for the longitudinal sector, and,
\begin{widetext}
\begin{eqnarray}
y_{1}^{2}&=&\frac{1}{2}\left(\mu^{2}+m^{2}+\Pi_{T}-\sqrt{-8Ng^{2}\gamma^{4}+\mu^{4}-2\mu^{2}m^{2}+m^{4}-2\mu^{2}\Pi_{T}+2m^{2}\Pi_{T}+\Pi_{T}^{2}}\right),\nonumber\\
y_{2}^{2}&=&\frac{1}{2}\left(\mu^{2}+m^{2}+\Pi_{T}+\sqrt{-8Ng^{2}\gamma^{4}+\mu^{4}-2\mu^{2}m^{2}+m^{4}-2\mu^{2}\Pi_{T}+2m^{2}\Pi_{T}+\Pi_{T}^{2}}\right),
\end{eqnarray}
\end{widetext}
for the transverse one.
This is of particular importance as it, in theory, precludes us from incorporating the conventional effective term $\frac{M^{2}}{2}A_{\mu}^{a}A_{\mu}^{a}$ into the action for describing the theory at finite temperatures. 

It is widely recognized that the condensates of the $2$-dimensional operators $A_{\mu}^{a}A_{\mu}^{a}$ and $\bar{\phi}^{ab}_{\mu}\phi^{ab}_{\mu}-\bar{\omega}^{ab}_{\mu}\omega^{ab}_{\mu}$ are exclusive to the infrared regime, implying that they may disappear as we approach the ultraviolet regime.

Given that each of these masses is determined by a gap equation, their resolution at finite temperatures would be necessary for analyzing their temperature dependence. However, as this falls outside the scope of our current work, we will defer such analysis to future research.

\section{Conclusions and future perspectives}
In this study, we examined the Refined Gribov-Zwanziger model at finite temperature to investigate the influence of the Gribov parameter and the masses of the condensates on the pole structure of the theory's propagators and their behavior. Our objective was to gain insights into the phase transitions of the theory resulting from variations in the temperature, denoted as $T$.

To do the analysis, a semi-classical technique was employed, wherein an effective massive Yang-Mills theory was examined to incorporate thermal effects into the poles of the gluon propagator. In essence, our approach involved the inclusion of a term $\frac{1}{2}M^{2}A_{\mu}A_{\mu}$ within the action of the theoretical framework. The term $M^{2}$ in this context denotes the comprehensive $1$-loop finite-temperature corrections of the theory, encompassing all pertinent thermal information. In the absence of this term, the observation of any form of phase transition to deconfined or semi-confined phase would be precluded.
Therefore, it is the thermal mass that plays a crucial role in the generation of a deconfined regime. 

Using this approach, we successfully determined two critical temperatures associated with phase transitions. This pertains to the transition from a fully constrained phase to a partially constrained one, involving both a tangible degree of freedom and an intangible one. The second phenomenon pertains to a phase transition occurring between the state of partial confinement and the state of total deconfinement.

All of these results are in complete accordance with the findings previously documented by Canfora in the Landau gauge \cite{canfora2014gribov}. In this study, we explore the broader scope of gauge considerations, specifically focusing on the Linear covariant gauges. Additionally, we investigate the impact of the masses of the RGZ condensates. It is evident from the analysis that their impact mostly entails the reduction of phase transition temperatures within the theoretical framework. 

Using this method, we were able to pinpoint two critical temperatures associated with phase transitions. The first is related to the transition from a purely confined phase to a partially confined one, characterized by a physical and a non-physical degree of freedom. The second corresponds to a phase transition from the partially confined phase to a completely deconfined one.

These findings align perfectly with those previously reported by Canfora in the Landau gauge in \cite{canfora2014gribov}. However, our current study extends beyond this by considering not only a more general gauge, namely the Linear covariant gauges, but also the impact of the masses of the RGZ condensates. Our observations indicate that these masses essentially lower the temperatures of phase transition in the theory.

As in \cite{canfora2014gribov}, we want to highlight that the emergence of an intermediate phase of partial confinement is a key finding of this study. This is particularly noteworthy as it aligns with the outcomes of other research \cite{liao2007strongly,pisarski2009towards,hidaka2008suppression,pisarski2013quasi}, where similar transition behaviour is exhibited by the quark-gluon plasma.

Upon concluding our work, we computed the thermal mass $M^{2}$ in relation to the Gribov parameter and the condensate masses. Our findings indicate that there are still temperature regimes where confining and deconfining phases exist. However, due to the complexity of the results, we were unable to formulate an effective action that could yield results mirroring those from eq. (\ref{prop finite t}). This is an aspect we aim to address in future work. In such a study, it would be intriguing to calculate the condensate masses by solving their respective gap equations at finite temperature. This would enable us to analyse the longitudinal and transverse masses $\Pi_{L}$ and $\Pi_{T}$ solely in terms of the temperature $T$, greatly facilitating our examination of the theory's phase transitions.

\section*{Acknowledgments}
The author Ferreira, L.C. acknowledges CNPq (Conselho Nacional de Desenvolvimento Científico e Tecnológico) for financial support under the grant PDJ (174040/2023-7).

\bibliography{refs}

\end{document}